\documentclass[manuscript,screen]{acmart}

\usepackage{threeparttable}
\usepackage{array}
\usepackage{multirow}
\usepackage{makecell}
\usepackage{bm}
\usepackage{enumitem}
\usepackage{blindtext}
\usepackage{hyperref}

\newcommand{\revision}[1]{\textcolor{black}{#1}}

\AtBeginDocument{%
  \providecommand\BibTeX{{%
    \normalfont B\kern-0.5em{\scshape i\kern-0.25em b}\kern-0.8em\TeX}}}

\setcopyright{acmcopyright}
\copyrightyear{2021}
\acmYear{2021}
\acmDOI{10.1145/3465371}

\acmJournal{JETC}

\begin{document}

\title{The Bitlet Model\\\normalsize A Parameterized Analytical Model to Compare PIM and CPU Systems}


\author{Ronny Ronen}
\affiliation{%
  \institution{Technion - Israel Institute of Technology}
  \city{Haifa}
  \country{Israel}
}

\author{Adi Eliahu}
\affiliation{%
  \institution{Technion - Israel Institute of Technology}
  \city{Haifa}
  \country{Israel}
}

\author{Orian Leitersdorf}
\affiliation{%
  \institution{Technion - Israel Institute of Technology}
  \city{Haifa}
  \country{Israel}
}

\author{Natan Peled}
\affiliation{%
  \institution{Technion - Israel Institute of Technology}
  \city{Haifa}
  \country{Israel}
}

\author{Kunal Korgaonkar}
\affiliation{%
  \institution{Technion - Israel Institute of Technology}
  \city{Haifa}
  \country{Israel}}

\author{Anupam Chattopadhyay}
\affiliation{%
 \institution{Nanyang Technological University}
 \city{Singapore}
 \country{Singapore}}

\author{Ben Perach}
\affiliation{%
  \institution{Technion - Israel Institute of Technology}
  \city{Haifa}
  \country{Israel}
}

\author{Shahar Kvatinsky}
\affiliation{%
  \institution{Technion - Israel Institute of Technology}
  \city{Haifa}
  \country{Israel}}

\renewcommand{\shortauthors}{Ronen, et al.}
\thanks{This work was supported by the European Research Council through the European Union’s Horizon 2020 Research and Innovation Programme under Grant 757259 and by the Israel Science Foundation under Grant 1514/17.}
\begin{abstract}
Nowadays, data-intensive applications are gaining popularity and, together with this trend, processing-in-memory (PIM)-based systems are being given more attention and have become more relevant. This paper describes an analytical modeling tool called Bitlet that can be used, in a parameterized fashion, to estimate the performance and the power/energy of a PIM-based system and thereby assess the affinity of workloads for PIM as opposed to traditional computing. The tool uncovers interesting tradeoffs between, mainly, the PIM computation complexity (cycles required to perform a computation through PIM), the amount of memory used for PIM, the system memory bandwidth, and the data transfer size. Despite its simplicity, the model reveals new insights when applied to real-life examples. The model is demonstrated for several synthetic examples and then applied to explore the influence of different parameters on two systems - IMAGING and FloatPIM. Based on the demonstrations, insights about PIM and its combination with CPU are concluded.

\end{abstract}



\maketitle

\section{Introduction}
\label{sec:Introduction}


Processing vast amounts of data on traditional von Neumann architectures involves many data transfers between the central processing unit (CPU) and the memory. These transfers degrade performance and consume energy~\cite{Ranganathan2011,  Seshadri2014, Seshadri2017, Fujiki2018, Eckert2018, IRAM}. Enabled by emerging memory technologies, recent memristive processing-in-memory (PIM)\footnote{We refer to memristive stateful logic~\cite{JohnPatmos} as PIM, but the concepts and model may apply to other technologies as well.} solutions show great potential in reducing costly data transfers by performing computations using individual memory cells~\cite{Raoux2008, IMPLYNature, Wong2012, Linn2012, Kvatinsky2014_1}. Research in this area has led to better circuits and micro-architectures~\cite{Kvatinsky2014_1, Kvatinsky2014_2, Bhattacharjee2017}, as well as applications using this paradigm~\cite{Imani2017, Haj2018}. 

PIM solutions have recently been integrated into application-specific~\cite{acceleratorSurvey} and general-purpose~\cite{Hur2016} architectures. General-purpose PIM-based architectures usually rely on memristive logic gates which are functionally complete sets to enable the execution of arbitrary logic functions within the memory. Different memristive logic techniques have been designed and implemented, including MAGIC~\cite{ Kvatinsky2014_1}, IMPLY~\cite{IMPLYNature}, resistive majority~\cite{MIG}, \revision{Fast Boolean Logic Circuit (FBLC,~\cite{FBLC}), and Liquid Silicon (\cite{LiquidSilicon})}. 

Despite the recent resurgence of PIM, it is still very challenging to analyze and quantify the advantages or disadvantages of PIM solutions over other computing paradigms. We believe that a useful analytical modeling tool for PIM can play a crucial role in addressing this challenge. An analytical tool in this context has many potential uses, such as in (i) evaluation of applications mapped to PIM, (ii) comparison of PIM versus traditional architectures, and (iii) analysis of the implications of new memory technology trends on PIM. 

Our Bitlet model (following~\cite{korgaonkar2019bitlet}) is an analytical modeling tool that facilitates comparisons of PIM versus traditional CPU\footnote{\revision{The Bitlet model concept can support systems other than CPU, \textit{e.g.}, GPU. See \textit{Comparing PIM to systems other than CPU} in Section \ref{sec:Model_Limitations}.}} computing. The name Bitlet reflects PIM's unique bit-by-bit data element processing approach. The model is inspired by past successful analytical models for computing~\cite{Gustafson1988, Hill2008, Williams2009, Esmaeilzadeh2011, Hill2019} and provides a simple operational view of PIM computations.  

The main contributions of this work are:  
\vspace{-3pt}
\begin{itemize} 
\item Presentation of use cases where using PIM has the potential to improve system performance by reducing data transfer in the system, and quantification of the potential gain and the PIM computation cost of these use cases.
\item Presentation of the Bitlet model, an analytical modeling tool that abstracts algorithmic, technological, as well as architectural machine parameters for PIM.
\item 
Application of the Bitlet model on various workloads to illustrate how it can serve as a \textit{litmus test} for workloads to assess their affinity on PIM as compared to the CPU.
\item Delineation of the strengths and weaknesses of the new PIM paradigm as observed in a sensitivity study evaluating PIM performance and efficiency over various Bitlet model parameters. 
\end{itemize}
\vspace{-3pt}

\revision{It should be emphasized that the Bitlet model is an exploration tool. Bitlet is intended to be used as an analysis tool for performing limit studies, conducting first-order comparisons of PIM and CPU systems, and researching the interplay among various parameters. Bitlet is \emph{not} a simulator for a specific system.}

The rest of the paper is organized as follows: Section~\ref{sec:Background} provides background on PIM. In Section~\ref{sec:Pim_Use_Cases}, we describe the PIM potential use cases. In Section \ref{sec:Performance}, we assess the performance of a PIM, CPU, and a PIM-CPU hybrid system. Section~\ref{sec:Power_and_Energy} discusses and compares the power and energy aspects of these systems. 
\revision{Note that Sections~\ref{sec:Pim_Use_Cases}-\ref{sec:Power_and_Energy} combine tutorial and research. These sections go deep into explaining step by step, using examples, both the terminology and the math behind PIM related use cases, performance, and power.} In Section~\ref{sec:Bitlet}, we present the Bitlet model and its ability to evaluate the potential of PIM and its applications. We conclude the paper in Section~\ref{sec:conclusion}.

\section{Background}
\label{sec:Background}

This section establishes the context of the Bitlet research. It provides information about current PIM developments, focusing on stateful logic-based PIM systems and outlining different methods that use stateful logic for logic execution within a memristive crossbar array.

\subsection{Processing-In-Memory (PIM)}
\label{sec:PIM}
The majority of modern computer systems use the von Neumann architecture, in which there is a complete separation between processing units and data storage units. Nowadays, both units have reached a scaling barrier, and the data processing performance is now limited mostly by the data transfer between them. The energy and delay associated with this data transfer are estimated to be several orders of magnitude higher than the cost of the computation itself~\cite{Pedram, IRAM}, and are even higher in data-intensive applications, which have become popular, \textit{e.g.}, neural networks~\cite{ISAAC} and DNA sequencing~\cite{GRIM_Filter}. This data transfer bottleneck is known as the \textit{memory wall}.

The memory wall has raised the need to bridge the gap between where data resides and where it is processed. First, an approach called \textit{processing-near-memory} was suggested, in which, computing units are placed close to or in the memory chip. Many architectures were designed using this method, \textit{e.g.}, intelligent RAM (IRAM)~\cite{IRAM}, active pages~\cite{ActivePages}, and 3D-stacked dynamic random access memory (DRAM) architectures~\cite{tesseract}. However, this technique still requires data transfer between the memory cells and the computing units. Then, another approach, called PIM was suggested, in which, the memory cells also function as computation units. Various new and emerging memory technologies, \textit{e.g.}, resistive random access memory (RRAM)~\cite{RRAM}, often referred to as memristors, have recently been explored. Memristors are new electrical components that can store two resistance values: $R_{ON}$ and $R_{OFF}$, and therefore can function as memory elements. In addition, by applying voltage or passing current through memristors, they can change their resistance and therefore can also function as computation elements. These two characteristics make the memristor an attractive candidate for PIM.

\subsection{Memristive Memory Architecture}
\label{sec:Memristive_Memory_architecture}
Like other memory technologies, memristive memory is usually organized in a hierarchical structure. Each RRAM chip is divided into \textit{banks}. Each bank is comprised of \textit{subarrays}, which are divided into two-dimensional memristive \textit{crossbars} (\textit{a.k.a.} \textit{XBs}). The XB consists of rows (\textit{wordlines}) and columns (\textit{bitlines}), with a memristive cell residing at each junction and logic performed within the XB. Overall, the RRAM chip consists of many XBs, which can either share the same controller and perform similar calculations on different data, or have separate controllers for different groups of XBs and act independently.

\subsection{Stateful Logic}
\label{sec:Stateful_Logic}
Different logic families, which use memristive memory cells as building blocks to construct logic gates within the memory array, have been proposed in the literature. These families have been classified into various categories according to their characteristics: statefulness, proximity of computation, and flexibility~\cite{JohnPatmos}. In this paper, we focus on ‘stateful logic' families, so we use the term PIM to refer specifically to stateful logic-based PIM, and we use the term \textit{PIM technologies} to refer to different stateful logic families. A logic family is said to be stateful if the inputs and outputs of the logic gates in the family are represented by memristor resistance. 

\begin{figure}[t]
 \begin{center}
  \includegraphics[width=3in]{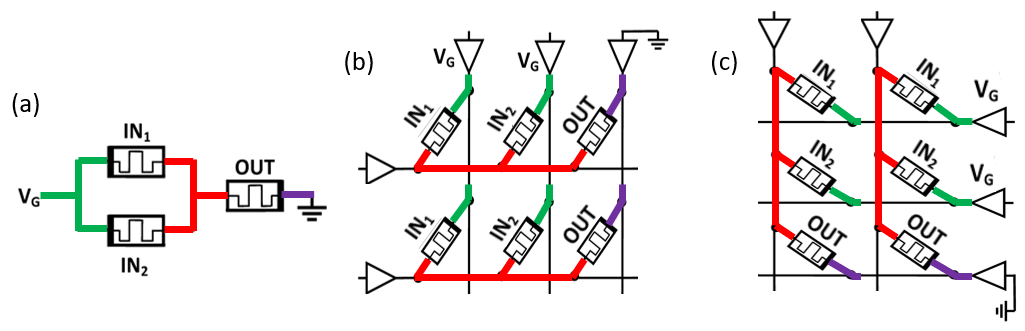}
 
 \caption{{MAGIC NOR gates. (a) MAGIC NOR gate schematic. (b) Two MAGIC NOR gates mapped to crossbar array rows, operated in parallel. (c) Two MAGIC NOR gates mapped to crossbar array columns, operated in parallel.}}
  \label{fig:MAGIC}
   \end{center} 
\vspace{-3mm}
\end{figure}

Several PIM technologies have been designed, including IMPLY~\cite{IMPLYNature} and MAGIC~\cite{Kvatinsky2014_1} gates. MAGIC gates have become a commonly used PIM technology. Figure~\ref{fig:MAGIC}(a) shows the MAGIC NOR logic gate structure, where the two input memristors are connected to an operating voltage, $V_{g}$, and the output memristor is grounded. Since MAGIC is a stateful logic family, the gate inputs and output are represented as memristor resistance. The input memristors are set with the input values of the logic gate and the output memristor is initialized at $R_{ON}$. The resistance of the output memristor changes during the execution according to the voltage divider rule, and switches when the voltage across it is higher than \(\frac{V_{g}}{2}\). The same gate structure can be used to implement an OR logic gate, with minor modifications (the output memristor is initialized at $R_{OFF}$ and a negative operating voltage $V_{g}$ is applied)~\cite{Barak_MAGIC_VCM}. As depicted in Figures~\ref{fig:MAGIC}(b) and \ref{fig:MAGIC}(c), a single MAGIC NOR gate can be mapped to a memristive crossbar array row (horizontal operation) or column (vertical operation). Multiple MAGIC NOR gates can operate on different rows or columns concurrently, thus enabling massive parallelism. Overall, logic is performed using the exact same devices that store the data.

\subsection{Logic Execution within a Memristive Crossbar Array}
\label{sec:Logic_Execution}
A functionally complete memristive logic gate, \textit{e.g.}, a MAGIC NOR gate, enables in-memory execution of any logic function. The in-memory execution is performed by a sequence of operations performed over several clock cycles. In each clock cycle, one operation can be performed on a single row or column, or on multiple rows or columns concurrently, if the data is row-aligned or column-aligned. The execution of an arbitrary logic function with stateful logic has been widely explored in the literature~\cite{SAID, SIMPLE, SIMPLER, YADAV}. Many execution and mapping techniques first use a synthesis tool, which synthesizes the logic function and creates a netlist of logic gates. Then, each logic gate in the netlist is mapped to several cells in the memristive crossbar and operated in a specific clock cycle. Each technique maps the logic function according to its algorithm, based on different considerations, \textit{e.g.}, latency, area, or throughput optimization.

Many techniques use several rows or columns in the memristive crossbar array for the mapping~\cite{SIMPLE,SAID,ReVAMP_Anupam} to reduce the number of clock cycles per a single function or to allow mapping of functions that are longer than the array row size \revision{by spreading them over several rows}. 
The unique characteristic of the crossbar array, which enables parallel execution of several logic gates in different rows or columns, combined with an efficient cell reuse feature that enables condensing long functions into short crossbar rows, renders single instruction multiple data (SIMD) operations attractive. In SIMD operations, the same function is executed simultaneously on multiple rows or columns. Executing logic in SIMD mode increases the computation throughput; therefore, by limiting the entire function mapping to a single row or column, the throughput can be substantially improved. This is applied in the SIMPLER~\cite{SIMPLER} mapper. \revision{Specifically, efficient cell reuse is implemented in SIMPLER by overwriting a cell when its old value is no longer needed. With cell reuse, SIMPLER can squeeze functions that require a long sequence of gates into short memory rows, \textit{e.g.}, a 128-bit addition that takes about 1800 memory cells without cell reuse is compacted into less than 400 memory cells with cell reuse.} In this paper, we assume, without loss of generality, that a logic function is mapped into a single row in the memristive crossbar and cloned to different rows for different data.

\subsection{\revision{The memristive Memory Processor Unit (mMPU) Architecture}}
\label{sec:mMPU}
\begin{figure}[t]
 \begin{center}
  \includegraphics[width=5in]{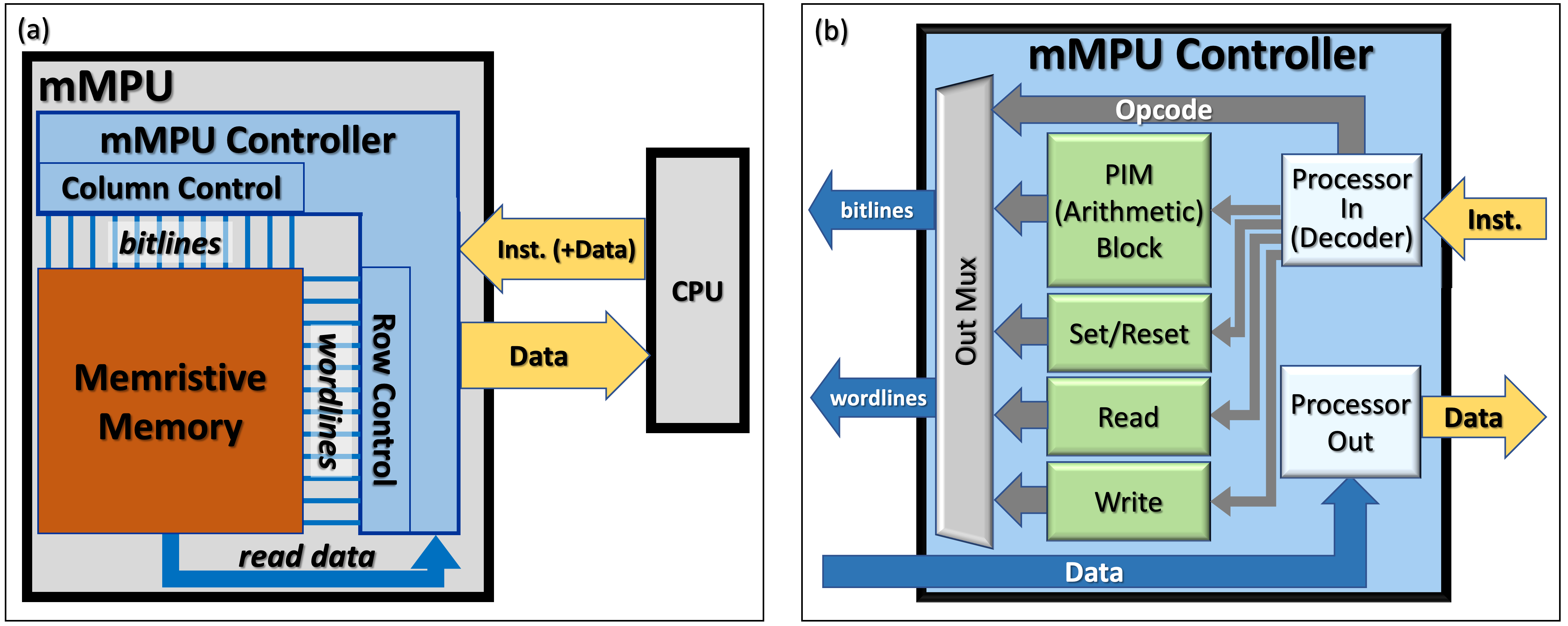}
  \caption{{\revision{The mMPU architecture. (a) The interfaces between the mMPU controller, the memory and the CPU. (b) A block diagram of the mMPU controller. The mMPU controller receives instructions from the CPU, optionally including data to be written, and returns read data to the CPU via the data lines.}}}
  \label{fig:mMPU}
   \end{center} 
   \vspace{-2mm}
\end{figure}

\revision{A PIM system requires a controller to manage its operation. In~\cite{controllerRotem}, a design of a memristive memory processing unit (mMPU) controller is presented. Figure~\ref{fig:mMPU} depicts the mMPU architecture, as detailed in~\cite{controllerRotem}.
Figure~\ref{fig:mMPU}(a) describes the interfaces between the mMPU controller, the memory, and the CPU. The CPU sends instructions to the mMPU controller, optionally including data to be written to memory. The mMPU processes each instruction and converts it into one or more memory commands. For each command, the mMPU determines the voltages applied on the wordlines and bitlines of the memristive memory arrays so that the command will be executed. Figure~\ref{fig:mMPU}(b) depicts the internal structure of the mMPU controller. The instruction is interpreted by the decoder and then further processed by one of the green processing blocks according to the instruction type. The set of mMPU instruction types consists of the traditional memory instruction types: Load, Store, Set/Reset, and a new family of instructions: PIM instructions. Load and store instructions are processed in the read and write blocks, respectively. Initialization instructions are processed in the Set/Reset block. PIM instructions are processed in the PIM ("arithmetic") block. The PIM block breaks each PIM instruction into a sequence of micro-instructions, and executes this sequence in consecutive clock cycles, as described in abstractPIM~\cite{abstractPIM}. The micro-instructions supported by the target mMPU controller can be easily adapted and modified according to the PIM technology in use, \textit{e.g.}, MAGIC NOR and IMPLY.
Load instructions return data to the CPU through the controller via the data lines.}

\revision{For PIM-relevant workloads, the overhead of the mMPU controller on the latency, power, and energy of the system is rather low. This is due to the fact that each PIM instruction operates on many data elements in parallel. Thus, the controller/instruction overhead is negligible relative to the latency, power, and energy cost of the data being processed within the memristive memory and the data transfer between memory and CPU~\cite{CONECPT}.}

\section{PIM Use Cases and Computation Principles}
\label{sec:Pim_Use_Cases}

After presenting the motivation for PIM in the previous section, in this section, we describe potential use cases of PIM. We start with a high-level estimate of the potential benefits of reduced data transfer. Later, we define some computation principles, and using them, we assess the performance cost of PIM computing. 


\subsection{PIM Use Cases and Data Transfer Reduction} 
\label{sec:PIM_Data_Reduction} 

As stated in Section~\ref{sec:Background}, the benefit of PIM comes mainly from the reduction in the amount of the data transferred between the memory and the CPU. If the saved time and energy due to the data transfer reduction is higher than the added cost of PIM processing, then PIM is beneficial. In this sub-section, we list PIM use cases that reduce data transfer and quantify this reduction.

For illustration, assume our data reflect a structured database in the memory that consists of $N$ records, where each record is mapped into a single row in the memory. Each record consists of fields of varying data types. A certain compute task reads certain fields of each record, with an overall size of $S_i$ bits, and writes back $S_o$ (potentially zero) bits as the output. Define $S = S_i + S_o$ as the total number of accessed bits per record. Traditional CPU-based computation consists of transferring $N \times S_i$ bits from memory to the CPU, performing the needed computations, and writing back $N \times S_o$ bits to the memory. In total, these computations require the transfer of $N \times S$ bits between the memory and the CPU. By performing all or part of the computations in memory, the total amount of data transfer can be reduced. This reduction is achieved by either reducing (or eliminating) the bits transferred per record ($"Compacting"$), and/or by reducing the number of transferred records ($"Filtering"$). 

\begin{figure}[t]
 \begin{center}
 \includegraphics[width=5in]{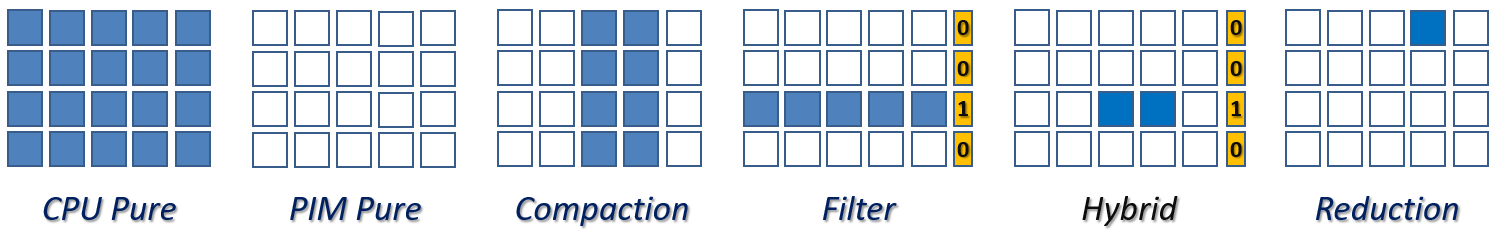}
 \caption{\small {Data size reduction illustration. Blue squares: data to be transferred, white: saved data transfer, yellow: bit vector of selected records to transfer.}}
  \label{fig:Data_Reduction}
 \end{center} 
\vspace{-3mm}
\end{figure}

Several potential use cases follow, all of which differ in the way a task is split between PIM and CPU. In all cases, we assume that all records are appropriately aligned in the memory, so that PIM can perform the basic computations on all records concurrently (handling unaligned data is discussed later in Section~\ref{sec:PIM_Computation_Principles}). Figure~\ref{fig:Data_Reduction} illustrates these use cases. 

\begin{itemize} 
\item \textbf{CPU Pure.} This is the baseline use case. No PIM is performed. All input and output data are transferred to the CPU and back. The amount of data transferred is $N \times S$ bits.  

\item \textbf{PIM Pure.} In this extreme case, the entire computation is done in memory and no data is transferred. This kind of computation is done, for example, as a pre-processing stage in anticipation of future queries. See relevant examples under \textit{PIM Compact} and \textit{PIM filter} below.     

\item \textbf{PIM Compact.} Each record is \revision{pre-}processed in memory in order to reduce the number of bits to be transferred to the CPU from each record. For example, each product record in a warehouse database contains 12 monthly shipment quantity fields. The application only needs the yearly quantity. Summing ("Compacting") these 12 elements into one reduces the amount of data transferred by 11 elements per record. Another example is an application that does not require the explicit shipping weight values recorded in the database, but just a short class tag (light, medium, heavy) instead. If the per-record amount of data is reduced from $S$ to $S_1$ bits, then the overall reduction is $N \times (S - S_{1})$ bits.

\begin{table}[t!]
\caption{PIM Use Cases Data Transfer Reduction} 
\label{tab:Use_Cases}
\vspace{-3mm}
\centering
\begin{threeparttable}
 \begin{tabular}{|c|c|c|c|c|}
 \hline
 Use Case & Records & Size & Data Transferred & Data Transfer Reduction \\
 \hline\hline
 $CPU$ $Pure$ & $N$ & $S$ &  $N \times S$ & $0$ \\
 \hline
 $PIM$ $Pure$ & $0$ & $0$ & $0$ & $N \times S$\\
 \hline
 $PIM$ $Compact$ & $N$ &  $S_1$ & $N \times S_1$ & $N \times (S - S_1)$\\
 \hline
 $PIM$ $Filter_1$ & $N_1$ &  $S$ & $N_1 \times S + N$ & $N \times S_1 - N$\\
 \hline
  $PIM$ $Filter_2$ & $N_1$ &  $S$ & $N_1 \times (S+log_2(N))$ & $N \times S - N_1 \times (S+log_2(N))$\\
 \hline
 $PIM$ $Hybrid$ & $N_1$ &  $S_1$ & $N_1 \times S_1+N$ & $N \times (S-1) - N_1 \times S_1$\\
 \hline
 $PIM$ $Reduction_0$ & $1$ &  $S_1$ & $1 \times S_1$ & $N \times S - S_1$\\
 \hline
 $PIM$ $Reduction_1$ & $N_1 = \lceil\frac{N}{R}\rceil$ &  $S_1$ & $N_1 \times S_1$ & $N \times S - N_1 \times S_1$\\
 \hline
\end{tabular}
\begin{tablenotes}[flushleft]
\setlength\tabcolsep{0pt}
\item $N$/$N_1$: overall/selected number of records; $S$/$S_1$: original/final size of records.
\item $Filter_1/Filter_2$: bit-vector/list of indices; $Reduction_0/Reduction_1$: all records/per XB
\end{tablenotes}
\end{threeparttable}
\vspace{-8pt}
\end{table}


\item \textbf{PIM Filter.} Each record is processed in memory to reduce the number of records transferred to the CPU. This is a classical database query case. For example, an application looks for all shipments over \$1M. Instead of passing all records to the CPU and checking the condition in the CPU, the check is done in memory and only the records that pass the check ("Filtering") are transferred. If only $N_1$ out of $N$ records of size $S$ are selected, then the overall data transfer is reduced by $(N - N_1) \times S$ bits. 
Looking deeper, we need to take two more factors into account:
\smallskip

\noindent (1) When the PIM does the filtering, the location of the selected records should also be transferred to the CPU, and the cost of transferring this information should be accounted for. Transferring the location can be done by  either ($Filter_1$) passing a bit vector ($N$ bits) or by ($Filter_2$) passing a list of indices of the selected records ($N_1 \times log_2(N)$ bits). The amount of the total data to be transferred is therefore $min(N \times 1, N_1 \times log_2(N))$. For simplicity, in this paper, we assume passing a bit vector ($Filter_1$). The overall cost of transferring both the data and the bit vector is $N_1 \times S+N$. The amount of saved data transfer relative to \textit{CPU Pure} is $N \times S_1 - N$ bits.
\smallskip

\noindent (2) When filtering is done on the CPU only, data may be transferred twice. First, only a subset of the fields (the size of which is $S_1$) that are needed for the selection process are transferred, and only then, the selected records or a different subset of the records. In this \textit{CPU Pure} case, the amount of transferred data is $N\times S_1 + N_1 \times S$.


\item \textbf{PIM Hybrid.} This use case is a simple combination of applying both \textit{PIM Compact} and \textit{PIM filter}. The amount of data transferred depends on the method we use to pass the list of selected records, denoted above as $Filter_1$ or $Filter_2$. For example, when using $Filter_1$, the transferred data consists of $N_1$ records of size $S_1$ and a bit-vector of size $N$ bits. That is $N_1 \times S_1+N$.

\item \textbf{PIM Reduction.} The reduction operator \textit{"reduces the elements of a vector into a single result"}\footnote{https://en.wikipedia.org/wiki/Reduction\_Operator}, \textit{e.g.}, computes the sum, or the minimum, or the maximum of a certain field in all records in the database. The size of the result may be equal to or larger than the original element size (\textit{e.g.}, summing a million of 8-bits arbitrary numbers requires 28 bits). "Textbook" reduction, referred to later as $Reduction_0$, replaces $N$ elements of size $S$ with a single element of size $S_1$ ($S_1 \geq S$), thus eliminating data transfer almost completely. A practical implementation, referred to later as $Reduction_1$, performs the reduction on each memory array (XB) separately and passes all interim reduction results to the CPU for final reduction. In this case, the amount of transferred data is the product of the number of memory arrays used by the element size, \textit{i.e.}, $\lceil\frac{N}{R} \rceil\times S_1$, where $R$ is the number of records (rows) in a single XB.
\end{itemize}

Table~\ref{tab:Use_Cases} summarizes all use cases along with the amount of transferred and saved data. In this table, $N$ and $N_1$ reflect the overall and selected number of the transferred records. $S$ and $S_1$ reflect the original and final size of the transferred records. 

\subsection{PIM Computation Principles} 
\label{sec:PIM_Computation_Principles} 

Stateful logic-based PIM (or just $PIM$ throughout this paper) computation provides very high parallelism. 
Assuming the structured database example above (Section \ref{sec:PIM_Data_Reduction}), where each record is mapped into a single memory row, PIM can generate only a single bit result per record per memory cycle (\textit{e.g.}, a single NOR, IMPLY, AND, based on the PIM technology). Thus, the sequence needed to carry out a certain computation may be rather long. Nevertheless, PIM can process many properly aligned records in parallel, \revision{computing one full column or full row per XB per cycle\footnote{\revision{The maximum size of a memory column or row may be limited in a specific PIM technology due to \textit{e.g.}, wire delays and write driver limitations.}}}. Proper alignment means that all the input cells and the output cell of all records occupy the same column in all participating memory rows (records), or, inversely, the input cells and the output cell occupy the same rows in all participating columns. 

PIM can perform the same operation on many independent rows, and many XBs, simultaneously. However, performing operations involving computations between rows (\textit{e.g.}, shift or reduction) or in-row copy of an element with a different alignment in each row, has limited parallelism. Such copies can be done in parallel among XBs, but within a XB, are performed mostly serially. When the data in a XB are aligned, operations can be done in parallel (as demonstrated in Figures~\ref{fig:MAGIC}(b) and~\ref{fig:MAGIC}(c) for row-aligned and column-aligned operations, respectively). However, operations on unaligned data cannot be done concurrently, as further elaborated in Section~\ref{sec:PIM_Computation_Principles}.

To quantify PIM performance, we first separate the computation task into two steps: \textit{Operation} and \textit{Placement and Alignment}. Below, we assess the complexity of each of these steps. For simplicity, we assume $N$ computations done on $R$ rows and $XBs$ memory arrays, \textit{i.e.}, $N = R \times XBs$.

\begin{figure}[t]
 \begin{center}
 \includegraphics[width=2.5in]{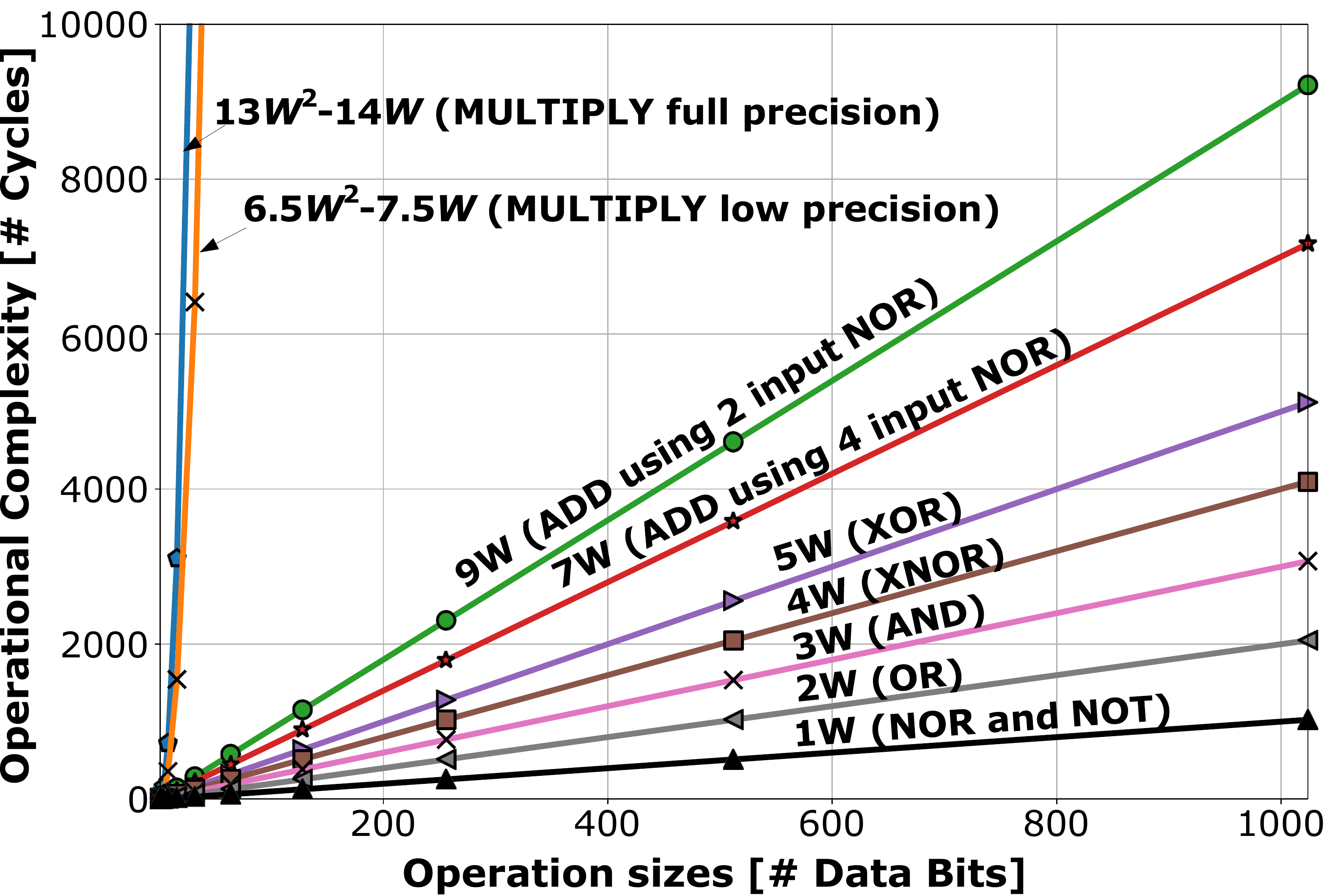}
 
 \caption{\small {PIM operation complexity in cycles for different types of operations and data sizes.}}
  \label{fig:funcs0}
 \end{center} 
\vspace{-3mm}
\end{figure}

\textbf{Operation Complexity ($OC$).} As described in Section~ \ref{sec:Stateful_Logic}, PIM computations are carried out as a series of basic operations, applied to the memory cells of a row inside a memristive memory array. 
While each row is processed bit-by-bit, the effective throughput of PIM is increased by the inherent parallelism achieved by simultaneous processing of multiple rows inside a memory array and multiple memory arrays in the system memory. We assume the same computations (\textit{i.e.}, individual operations) applied to a row are also applied in parallel in every cycle across all the rows ($R$) of a memory array. 

We define \textbf{Operation Complexity} ($OC$) for a given operation type and data size, as the number of cycles required to process the corresponding data. Figure~\ref{fig:funcs0} shows how the input data length ($W$) affects the computing cycles for PIM-based processing. The figure shows that this number is affected by both the data size, as well as operation types (different operations follow a different curve on the graph). In many cases, $OC$ is linear with the data size, for example, in a MAGIC NOR-based PIM, $W$-bit AND requires $3W$ cycles (\textit{e.g.}, for $W$=16 bits, AND takes 16x3 = 48 cycles), while ADD requires 9$W$ cycles\footnote{ADD can be improved to 7$w$ cycles using four-input NOR gates instead of two-input NOR gates.}. Some operations, however, are not linear, \textit{e.g.}, full precision MULTIPLY $W\times W\rightarrow 2W$ bits requires $13W^2-14W$ cycles~\cite{Haj2018} or approximately $12.5W^2$ cycles, while low precision MULTIPLY $W\times W\rightarrow W$ bits requires about half the number of cycles, or approximately $6.25W^2$ cycles. The specific Operation Complexity behavior depends on the PIM technology, but the principles are similar.

\textbf{Placement and Alignment Complexity} ($PAC$). PIM imposes certain constraints on data alignment and placement~\cite{Talati2018}. To align the data for subsequent row-parallel operations, a series of data alignment and placement steps, consisting of copying data from one place to another, may be needed. The number of cycles needed to perform these additional copy steps is captured by the placement and alignment complexity parameter, denoted as $PAC$. Currently, for simplicity, we consider only the cost of intra-XB data copying, we ignore the cost of inter-XB data copying, and we assume that multiple memory arrays continue to operate in parallel and independently. Refining the model to account for inter-XB data copying will be considered in the future (see Section~\ref{sec:Model_Limitations}).



\begin{figure}[t]
 \begin{center}
 \includegraphics[width=5in]{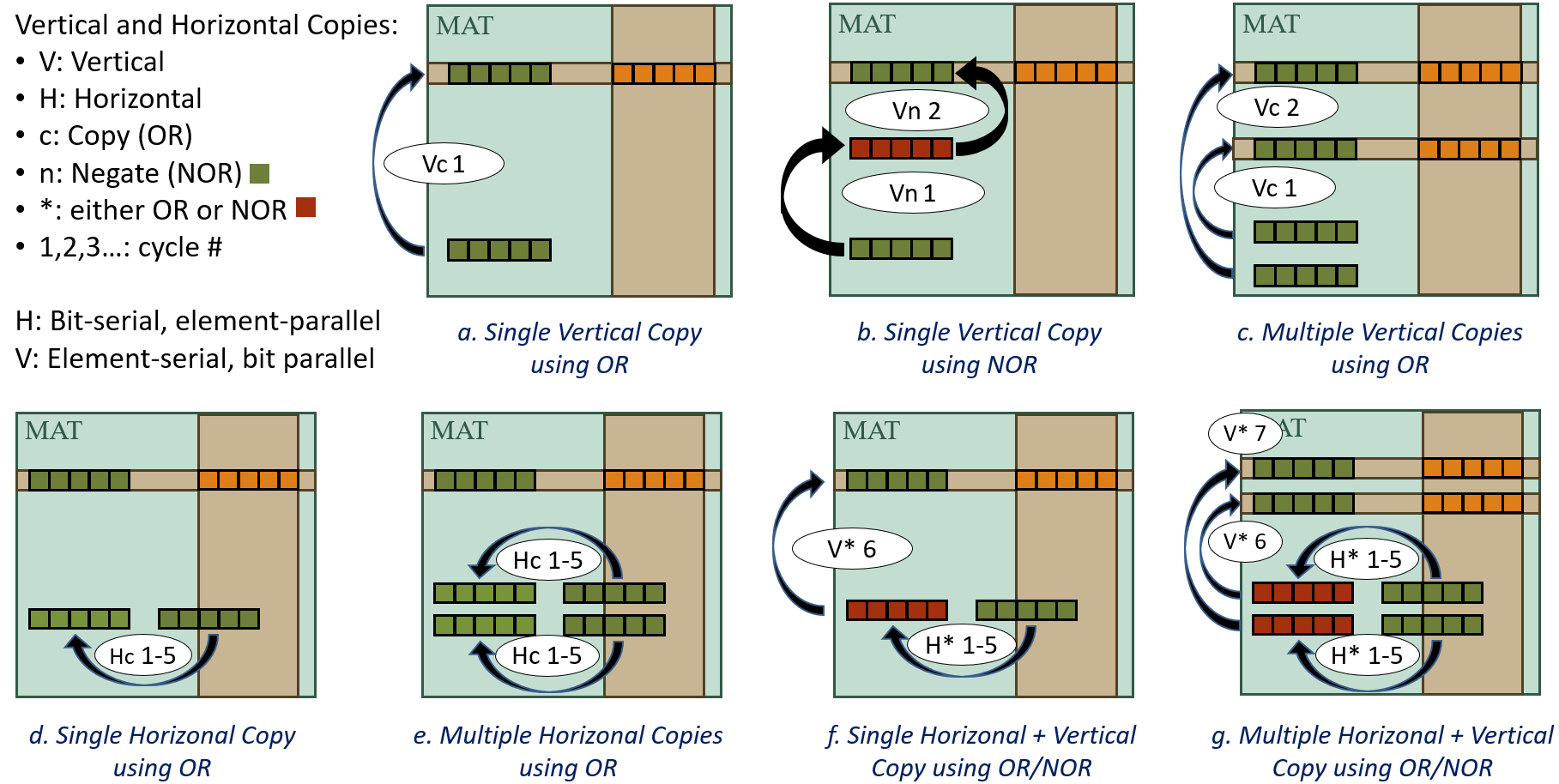}
 
 \caption{\small {Horizontal Copies ($HCOPY$) and Vertical Copies ($VCOPY$) using PIM. \revision{HCOPY: all elements move together, bit per cycle (\textit{case~e}). VCOPY: all bits move together, element per cycle (\textit{case~c}). Applied together in \textit{case~g}}.}}
  \label{fig:HV-COPY}
 \end{center} 
\vspace{-3mm}
\end{figure}

The PAC cycles required to copy the data in a memory array to the desired locations can be broken down into a series of horizontal row-parallel copies ($HCOPY$), and vertical column-parallel copies ($VCOPY$). 
\revision{Figure~\ref{fig:HV-COPY} shows examples of VCOPY and HCOPY operations involving copying a single element (Figures~\ref{fig:HV-COPY}(a), \ref{fig:HV-COPY}(b), \ref{fig:HV-COPY}(d)) and multiple elements (Figures~\ref{fig:HV-COPY}(c), \ref{fig:HV-COPY}(e), \ref{fig:HV-COPY}(f), \ref{fig:HV-COPY}(g)).
HCOPYs and VCOPYs are symmetric operations: HCOPY can copy an entire memory column (or part of it) in parallel, while VCOPY can copy an entire memory row  (or part of it) in parallel. 
Figure~\ref{fig:HV-COPY}(g) depicts the case of copying $m$ column-aligned elements, each $W$-bit wide (in green, $m$=2, $n$=5), into different rows to be placed in the same rows as other column aligned elements (in orange).
First, HCOPYs are performed in a \textit{bit-serial, element-parallel}, manner (copying elements from green to brown). In the first HCOPY cycle, all the first bits of all involved $m$ elements are copied in parallel. Then, in the second cycle, all the second bits of all involved $m$ elements are copied in parallel. This goes on for $W$ cycles until all $W$ bits in all $m$ elements are copied. Next, VCOPYs are performed in an \textit{element-serial, bit-parallel} manner (copying from brown to green). In the first VCOPY cycle, all the $W$ bits of the first selected element are copied, in parallel, to the target row. Then, in the second cycle, all the $W$ bits of the second selected element are copied, in parallel. This goes on for $m$ cycles until all $m$ elements are copied.}

When the involved data elements across different rows are not aligned, separate HCOPYs are performed individually for each data element, thus requiring additional cycles. A VCOPY for a given data element, on the other hand, can be done in parallel on all the bits in the element, which are in the same row. However, each row within a XB has to be vertically copied separately, in a serial manner.

The number of cycles it takes to perform a single bit copy (either HCOPY or VCOPY) depends on the PIM technology used. For example, MAGIC OR-based PIM technology (~\cite{Barak_MAGIC_VCM}) supports logic OR as a basic operation, allowing a 1-cycle bit copy (see Figure~\ref{fig:HV-COPY}(a),~\ref{fig:HV-COPY}(c),~\ref{fig:HV-COPY}(d), and~\ref{fig:HV-COPY}(e)). PIM Technologies that do not support a 1-cycle bit copy (\textit{e.g.}, MAGIC NOR-based PIM technology), have to execute two consecutive NOT operations that take two cycles to copy a single bit (Figure~\ref{fig:HV-COPY}(b)). However, copying a single bit using a sequence of a HCOPY operation followed by a VCOPY operation can be implemented as two consecutive OR or NOT operations that take two cycles regardless of the PIM technology used (Figure~\ref{fig:HV-COPY}(f) and~\ref{fig:HV-COPY}(g)).

We define \textbf{Computation complexity} ($CC$) as the number of cycles required to fully process the corresponding data. $CC$ equals the sum of $OC$ and $PAC$.

Below are examples of PIM $CC$ cycles. We use the terms \textit{\revision{Gathered}} and \textit{\revision{Scattered}} to refer to the \revision{original layout of the elements to be aligned}. \textit{\revision{Gathered}} means that all input locations are fully aligned among themselves, but not with their destination, while \textit{\revision{Scattered}} means that input locations are not aligned among themselves.
\vspace{-4pt}
\begin{itemize} 
\item \textbf{Parallel aligned operation.} Adding two vectors, $A$ and $B$, into vector $C$, where $A_i$, $B_i$ and $C_i$ are in row $i$. The size of each element is $W$-bits. A MAGIC NOR-based full adder operation takes $o (=9)$ cycles. Adding two $W$-bit elements in a single row takes $OC = o \times W$ cycles. At the same $OC$ cycles, one can add either one element or millions of elements. Since there are no vertical or horizontal copies, the $CC$ equals the $OC$. The above-mentioned PIM Compact, PIM Filter, and PIM Hybrid use cases are usually implemented as parallel aligned operations.

\item \textbf{\revision{Gathered} placement and alignment copies.} Assume we want to perform a shifted vector copy, \textit{i.e.}, copying vector $A$ into vector $B$ such that $B_{i-1} \leftarrow A_i$.\footnote{We ignore the elements of $B$ that are last in each XB.} The size of each element is $W$-bit. With stateful logic, the naive way of making such a copy for a single element is by a sequence of $HCOPY$ operations followed by $VCOPY$ operations. For a given single element in a row $A_i$, first, copy all $W$ bits of $A_i$ in parallel, so $B_i \leftarrow {A_i}$, then, copy $B_{i-1} \leftarrow {B_i}$. Copying $W$-bits in a single row
takes $W$ cycles. As in the above parallel aligned case, in the same $W$ cycles, one can copy either one element or many elements. However, in this case, we also need to copy the result elements from one row to the adjacent one above. Copying $W$-bits between two rows takes a single cycle, as all $W$ bits can be copied from one row to another in parallel. But, copying all rows is a serial operation, as it must be done separately for each row in the XB. Hence, if the memory array contains $R$ rows, the entire copy task will take $CC = (W + R)$ cycles. Still, these operations can be done in parallel on all the XBs in the system. Hence, copying all $N$ elements can be completed in the same $(W + R)$ cycles.

\item \textbf{\revision{Gathered} unaligned operation.} The time to perform a combination of the above two operations, \textit{e.g.}, $C_{i-1} \leftarrow A_i+B_i$, is the sum of both computations, that is $CC = (OC+W+R)$ cycles.

\item \textbf{\revision{Scattered} placement and alignment.} We want to gather \revision{$R$ unaligned $W$-bit} elements into a row-aligned vector $B$, that is, all $B$ elements occupy the same columns. Assume the worst case where all elements have to be horizontally and vertically copied to reach their desired location, as described above for \textit{\revision{Gathered} placement and alignment}. To accomplish this, we need to do $W$ horizontal 1-bit copies and one parallel $W$-bit copy for each element, totaling overall $CC = (W+1)\times R$ cycles.

\item \textbf{\revision{Scattered} unaligned operation.} Perform a \textit{\revision{Scattered} placement and alignment} followed by a \textit{parallel aligned operation}, takes the sum of both computations, that is, $CC = (OC+(W+1)\times R)$ cycles.

\item \textbf{Reduction.} We look at a classical reduction where the reduction operation is both commutative and associative (\textit{e.g.}, a sum, a minimum, or a maximum of a vector). For example, we want to sum a vector $A$ where each element, as well as the final sum, are of size $W$, \textit{i.e.}, $S = \sum_{i=1}^{N} A_i$. The idea is to first reduce all elements in each XB into a single value separately, but in parallel, and then perform the reduction on all interim results. There are several ways to perform a reduction, the efficiency of which depends on the number of elements and the actual operation. We use the tree-like reduction\footnote{https://en.wikipedia.org/wiki/Graph\_reduction}, which is a phased process, in which at the beginning of each phase, we start with $k$ elements ($k = R$, the number of rows, in the first phase), pair them into $k/2$ groups, perform all $k/2$ additions, and start a new phase with the $k/2$-generated numbers. For $R$ elements, we need $ph=\lceil \log_2(R) \rceil$ phases. Each phase consists of one parallel (horizontal) copy of $W$ bits, followed by $k/2$ serial (vertical) copies, and ending with one parallel operation (in our case, $W$-bit add).  The total number of vertical copies is $R-1$. Overall, the full reduction of a single XB in all phases takes $CC = (ph \times(OC+W)+(R-1))$ cycles. The reduction is done on all involved XBs in parallel, producing a single result per XB. Later, all per XB interim results are copied into fewer XBs and the process continues recursively over $\frac{\log_2(N)}{\log_2(R)}$ steps. Copying all interim results into fewer XBs and using PIM on a smaller number of XBs is inefficient as it involves serial inter-XB copies and low-parallel PIM computations. Therefore, for higher efficiency, after the first reduction step is done using PIM, all interim results are passed to the CPU for the final reduction, denoted as $Reduction_1$ in Section~\ref{sec:Pim_Use_Cases}.


\end{itemize}

\begin{table}[h!]
\centering
\caption{PIM Computation Cycles for Aligned and Unaligned Computations}
\vspace{-3mm}
\begin{threeparttable}
 \begin{tabular}{|>{\centering\arraybackslash}m{4.4cm}|>{\centering\arraybackslash}m{1cm}|>{\centering\arraybackslash}m{1cm}|>{\centering\arraybackslash}m{1cm}|>{\centering\arraybackslash}m{2.7cm}|c|}
 \hline
Computation type & Operate \textit{Row Parallel} & HCOPY \textit{Row Parallel} & VCOPY \textit{Row Serial} & Total & Approximation \\
 \hline\hline
 Parallel Operation & $OC$ & - & - & $OC$ & $OC$ \\
 \hline
\revision{Gathered} Placement \& Alignment & - & $W$ & $R$ & $W+R$ & $R$ \\
 \hline
\revision{Gathered} Unaligned Operation& $OC$ & $W$ & $R$ & $OC+W+R$ & $OC+R$ \\
 \hline
\revision{Scattered} Placement \& Alignment & - & $R\times W$ & $R$ & $(W+1) \times R$ & $W \times R$ \\
 \hline
\revision{Scattered} Unaligned Operation& $OC$ & $R\times W$ & $R$ & $OC+(W+1) \times R$ &  $OC+W \times R$ \\
 \hline
$Reduction_1$ & $ph \times OC$ & $ph \times W$ & $R-1$ & $ph \times(OC+W)+(R-1)$ & $ph \times OC+R$ \\
  \hline
\end{tabular}
\begin{tablenotes}[flushleft]
\setlength\tabcolsep{0pt}
\item $OC$: Operation Complexity, $W$: Width of element, $R$: Number of rows, $ph$: Number of reduction phases. 
\end{tablenotes}
\end{threeparttable}
\label{tab:PIMcycles}
\end{table}

Table~\ref{tab:PIMcycles} summarizes the computation complexity in cycles of various PIM computation types (ignoring inter-XB copies, as mentioned above). Usually, $OC\gg W$ and $R\gg W$, so $OC\pm W$ is approximately $OC$, and  $R \pm 1$ and $R\pm W$ are approximately $R$. The last column in the table reflects this approximation. The approximation column hints to where most cycles go, depending on $OC$, $R$, and $W$. Parallel operations depend on only $OC$ and are independent of $R$, the number of elements (rows). When placement and alignment take place, there is a serial part that depends on $R$ and is a potential cause for computation slowdown.

\section{PIM and CPU Performance} 
\label{sec:Performance} 

In the previous section, the PIM use cases and computation complexity were introduced. In this section, we devise the actual performance equations of PIM, CPU, and combined PIM+CPU systems.


\subsection{PIM Throughput} 
\label{sec:PIM_Throughput} 

$CC$ represents the time it takes to perform a certain computation, similar to the latency of an instruction in a computing system. However, due to the varying parallelism within the PIM system, $CC$ does not directly reflect the PIM system performance. To evaluate the performance of a PIM system, we need to find its system throughput, which is defined as the number of computations performed within a time unit. Common examples are Operations Per Second (OPS) or Giga Operations per Second (GOPS). For a PIM Pure case, when completing $N$ computations takes $T_{PIM}$ time, the PIM throughput  $TP_{PIM}$ is:
\begin{gather} 
{\textstyle  TP_{PIM}  =  \frac{N}{T_{PIM}}}. \label{eq:TPp} 
\end{gather}  

To determine $T_{PIM}$, we obtain the PIM $CC$ and multiply it by the PIM cycle time ($CT$)\footnote{\revision{We assume that the controller impact on the overall PIM latency is negligible, as explained in Section~\ref{sec:mMPU}.}}. $CT$ depends on the specific PIM technology used. To compute $CC$, we use the equations in Table \ref{tab:PIMcycles}. The number of computations $N$ is the total number of elements participating in the process. When single-row based computing is used, this number is the number of all participating rows, which is the product of the number of rows within a XB with the number of XBs, that is, $N = XBs\times R$.
The PIM throughput is therefore
\begin{gather} 
{\textstyle  TP_{PIM}  =  \frac{XBs \times R}{CC\times CT}}.
\label{eq:TPp1} 
\end{gather}  

For example, consider the \textit{\revision{Gathered} unaligned operation} case for computing shifted vector-add, $C_{i-1} \leftarrow A_i+B_i$. Assuming $o=9$ cycles ($1$-bit add), element size $W=16$ bits, $R=1024$ rows, and $XBs = 1024$ memory arrays, then $N=1M$ elements. $OC=9\times 16=144$ cycles. The number of cycles to compute $R$ elements also equals the time to compute $N$ elements and is $CC = OC+R$ or $144+512=656$ cycles. The PIM throughput per cycle is $\frac{N}{CC}=\frac{1024\times 1024}{656}=1598$ computations per cycle. The throughput is $\frac{1598}{CT}$ computations per time unit. We can derive the throughput per second for a specific cycle time. For example, for a $CT$ of 10ns, the PIM throughput is $\frac{1598}{10^{-8}}=159.8\times 10^9$ OPS $ \approx 160$ GOPS.

In the following sections, we explain the CPU Pure performance and throughput and delve deeper into the overall throughput computation when both PIM and CPU participate in a computation.

\subsection{CPU Computation and Throughput} 
\label{sec:CPU_Computations} 

Performing computation on the CPU involves moving data between the memory and the CPU (\textit{Data Transfer}), and performing the actual computations (\textit{e.g.}, ALU Operations) within the CPU core (\textit{CPU Core}). Usually, on the CPU side, Data Transfer and CPU Core Operations can overlap, so the overall CPU throughput $TP_{CPU}$ is the minimum between the data transfer throughput and the CPU Core Throughput.


Using PIM to accelerate a workload is only justified when the workload performance bottleneck is the data transfer between the memory and the CPU, rather than the CPU core operation. \revision{In such workloads, the data-set cannot fit in the cache as the data-set size is much larger than the CPU cache hierarchy size. Cases where the CPU core operation, rather than the data transfer, is the bottleneck, are not considered PIM-relevant.} In PIM-relevant workloads, the overall CPU throughput $TP_{CPU}$ is dominated by the data transfer throughput.

The data transfer throughput depends on the memory to CPU bandwidth and the amount of data transferred per computation. We define $BW$ as the memory to CPU bandwidth in bits per second (bps), and $DIO$ ($DATA~IO$) as the number of bits transferred for each computation. That is:
\begin{gather} 
{\textstyle  TP_{CPU}  =  \frac{BW}{DIO}}
\label{eq:TP_{CPU}}. 
\end{gather}  

We demonstrate the data transfer throughput using, again, the shifted 16-bit vector-add example. In Table \ref{tab:DataThroughput}, we present three interesting cases, differing in their $DIO$ size. (a) \textbf{CPU Pure.} The two inputs and the output are transferred between the memory and the CPU ($DIO=48)$. (b) \textbf{Inputs only.} Same as CPU Pure, except that only the inputs are transferred to the CPU; no output result is written back to memory ($DIO=32)$. (c) \textbf{Compaction.} Where PIM performs the add operation and passes only the output data to the CPU for further processing ($DIO=16)$. We use the same data bus bandwidth, $BW=1000$ GOPS, for all three cases. Note that the data transfer throughput depends only on the data sizes, it is independent of the operation type. The throughput numbers in the table reflect any binary 16-bit operation, either simple as OR or complex as divide. The table hints at the potential gain that PIM opens by reducing the amount of data transfer between the memory and CPU. If PIM throughput is sufficiently high, the data transfer reduction may compensate for the additional PIM computations and the combined PIM+CPU system throughput may exceed the throughput of a system using the CPU only with PIM.  

Special care must be taken when determining DIO for the PIM Filter and PIM Reduction cases since only a subset of the records are transferred to the CPU. Note that the DIO parameter reflects the number of data bits transferred per accomplished computation, even though the data for some computations were not eventually transferred. In these cases, the DIO should be set as the total number of transferred data bits divided by the number of computations done in the system. For example, assume a filter, where we process $N$ records of size $S$, and pass only $M=N\times p$ of them ($p<1$). The DIO, in case we use a bit-vector to identify chosen records, is $\frac{(N\times p \times S)+N}{N} = (S\times p) +1$. \textit{e.g.}, if $S=200$ and $p=1\%$, DIO is $200\times 0.01+1 = 2+1 = 3$ bits. That is, the amount of data transfer per computation went from 200 to 3 bits per computation, \textit{i.e.},  $67\times$ reduction. The data transfer throughput for the filter case is presented in Table \ref{tab:DataThroughput}.

\begin{table}[h!]
\centering
\caption{Data Transfer Throughput}
\vspace{-3mm}
\begin{tabular}{|>{\centering\arraybackslash}m{2.8cm}|>{\centering\arraybackslash}m{2.5cm}|>{\centering\arraybackslash}m{2cm}|>{\centering\arraybackslash}m{5cm}|}
 \hline
Computation type & Bandwidth (BW) [Gbps] & DataIO (DIO) [bits] & Data Transfer Throughput ($TP_{CPU}$) [GOPS] \\
 \hline\hline
 \textit{CPU Pure} & 1000 & 48 & 20.8 \\ 
 \hline
\textit{Inputs Only} & 1000 & 32 & 31.3 \\
 \hline
\textit{Compaction} & 1000 & 16 & 62.5 \\ 
 \hline
\textit{Filter (200 bit, 1\%)} & 1000 & 3 & 333.3 \\ 
 \hline
 \end{tabular}
\label{tab:DataThroughput}
\end{table}

\subsection{Combined PIM and CPU Throughput} 
\label{sec:Combined_Throughput}
\revision{In a combined PIM and CPU system, achieving peak PIM throughput requires operating all XBs in parallel, thus preventing overlapping PIM computations with data transfer\footnote{\revision{Overlapping PIM computation and data transfer can be made possible in a banked PIM system. See \textit{Pipelined PIM and CPU} in Section \ref{sec:Model_Limitations}.}}.}
In such a system, completing $N$ computations takes $T_{PIM}$ PIM time and $T_{CPU}$ data transfer time, and the combined throughput $TP_{Combined}$ is, by definition:
\begin{gather} 
{\textstyle  TP_{Combined}  =  \frac{N}{T_{PIM} + T_{CPU}}}.
\label{eq:TP_Combined0} 
\end{gather}  


Fortunately, computing the combined throughput $TP_{Combined}$ does not require knowing the values of $T_{PIM}$ and $T_{CPU}$.  $TP_{Combined}$ can be computed using the throughput values of its components, $TP_{PIM}$ and $TP_{CPU}$, as follows:
\begin{gather} 
 {\textstyle  TP_{Combined} = \frac{N}{T_{PIM} + T_{CPU}} = 
 \frac{1}{\frac{T_{PIM}}{N} + \frac{T_{CPU}}{N}} =  
 \frac{1}{\frac{1}{TP_{PIM}} + \frac{1}{TP_{CPU}}}}.
 \label{eq:TP_Combined} 
\end{gather}  

Since the PIM and CPU operations do not overlap, the combined throughput is always lower than the throughput of each component for the pure cases with the same parameters. For example, in the \textit{\revision{Gathered} unaligned operation} case above, when computing a 16-bit shifted vector-add, \textit{i.e.}, $C_{i-1} \leftarrow A_i+B_i$, we do the vector-add in PIM, and transfer the 16-bit result vector to the CPU (for additional processing). We have already shown that, for the parameters we use, the PIM Throughput is $TP_{PIM}=160$ GOPS, and the data transfer throughput is $TP_{CPU}=62.5$ GOPS. Using Eq.~(\ref{eq:TP_Combined}), the combined throughput $TP_{Combined}=\frac{1}{\frac{1}{160\times 10^9} + \frac{1}{62.5\times 10^9}} = 44.9\times 10^9=44.9$ GOPS, which is indeed lower than 160 and 62.5 GOPS. However, this combined throughput is higher than that of the CPU Pure throughput using higher $DIO$=32 or $DIO$=48 (31.3 or 20.8 GOPS) presented in the previous subsection. Of course, these results depend on the specific parameters used here. A comprehensive analysis of the performance sensitivity is described in Section~\ref{sec:Applying_Bitlet_Model}.

\section{Power and Energy} 
\label{sec:Power_and_Energy}

When evaluating the power and energy aspects of a system, we examine two factors:
\begin{itemize} 
\item \textbf{Energy per computation.} The energy needed to accomplish a single computation. This energy is determined by the amount of work to be done (\textit{e.g.}, number of basic operations) and the energy per operation. Different algorithms may produce different operation sequences thus affecting the amount of work to be done. Physical characteristics of the system affect the energy per operation. Energy per Computation is a measure of system efficiency. A system configuration that consumes less energy per a given computation is considered more efficient. For convenience, we generally use Energy Per Giga Computations. Energy is measured in Joules.
\item \textbf{Power.} The power consumed while performing a computation. The maximum allowed power is usually determined by physical constraints like power supply and thermal restrictions, and may limit system performance. It is worth noting that the high parallel computation of PIM causes the memory system to consume much more power when in PIM mode than when in standard memory load/store mode. Power is measured in Watts (Joules per second).
\end{itemize}

In this section, we evaluate the PIM, the CPU, and the combined system power and energy per computation and how it may impact system performance. For the sake of this coarse-grained analysis, we consider dynamic power only and ignore power management and dynamic voltage scaling.

\subsection{PIM Power and Energy} 
\label{sec:PIM_Power}

\revision{Most power models target a specific design. The below approach is more general and resembles the one used for floating-point add/multiply power estimation in FloatPIM~\cite{FloatPIM}. In this approach,}  
every PIM operation consumes energy. For simplicity, we assume that in every PIM cycle, the switching of a single cell consumes a fixed energy $Ebit_{PIM}$. \revision{This is the average amount of energy consumed by each participating bit in each XB, and accounts for both the memristor access as well as other overheads such as the energy consumed by the wires and the peripheral circuitry connected to the specific bitline/wordline\footnote{\revision{We assume that the controller impact on the overall PIM power and energy is negligible, as explained in Section~\ref{sec:mMPU}}.}}. The PIM energy per computation $EPC_{PIM}$ is the product of $Ebit_{PIM}$ by the number of cycles $CC$. The PIM power $P_{PIM}$ is the product of the energy per computation $EPC_{PIM}$ by the PIM throughput $TP_{PIM}$ (see Section~\ref{sec:PIM_Throughput}).
\begin{gather} 
 {\textstyle  
 EPC_{PIM} = Ebit_{PIM} \times CC},\\
 \textstyle P_{PIM} =  EPC_{PIM} \times TP_{PIM} = 
 (Ebit_{PIM} \times CC) \times 
 \frac{XBs \times R}{CC\times CT} =
 \frac {Ebit_{PIM} \times R\times XBs}{CT}.
  \label{eq:P_PIM} 
\end{gather}  

\subsection{CPU Power and Energy} 
\label{sec:CPU_Power}

Here we compute the CPU energy per computation $EPC_{CPU}$ and the power $P_{CPU}$. As in the performance model, we ignore the actual CPU Core operations and consider only the data transfer power and energy. Assume that transferring a single bit of data consumes $Ebit_{CPU}$. Hence, the CPU energy per computation $EPC_{CPU}$ is the product of $Ebit_{CPU}$ by the number of bits per computation $DIO$. The CPU power $P_{CPU}$ is simply the product of the energy per computation  $EPC_{CPU}$ with the CPU Throughput $TP_{CPU}$. When the memory to CPU bus is not idle, the CPU power $P_{CPU}$ is equal to the product of the energy per bit $Ebit_{CPU}$ with the number of bits per second, which is the memory to CPU bandwidth $BW$. 
\begin{gather} 
 {\textstyle EPC_{CPU} = Ebit_{CPU} \times DIO}, \\
 {\textstyle  P_{CPU} = EPC_{CPU} \times TP_{CPU} = 
  Ebit_{CPU} \times DIO \times \frac{BW}{DIO} = Ebit_{CPU} \times {BW}}
  \label{eq:P_CPU0}. 
\end{gather}  

If the bus is busy only part of the time, the CPU power $P_{CPU}$ should be multiplied by the relative time the bus is busy, that is, the bus duty cycle, 

\subsection{Combined PIM and CPU Power and Energy}
\label{sec:Combined_Power}

When a task is split between PIM and CPU, we treat them  as if part of each computation is partly done on the PIM and partly on the CPU (see Section~\ref{sec:Combined_Throughput}). The combined energy per computation $EPC_{Combined}$ is the sum of the PIM energy per computation $EPC_{PIM}$ and the CPU energy per computation $EPC_{CPU}$. The overall system power is the product of the combined energy per computation and the combined system throughput:
\begin{gather} 
 {\textstyle EPC_{Combined} = EPC_{PIM} + EPC_{CPU} = \frac{P_{PIM}}{TP_{PIM}} + \frac{P_{CPU}}{TP_{CPU}} },\\
 {\textstyle  P_{Combined} = EPC_{Combined} \times TP_{Combined} = 
 (\frac{P_{PIM}}{TP_{PIM}} + \frac{P_{CPU}}{TP_{CPU}}) \times TP_{Combined} }.
 \label{eq:P_CPU} 
\end{gather}  

Since PIM and CPU computations do not overlap, their duty cycle is less than 100\%. Therefore, the PIM power in the combined PIM+CPU system is lower than the maximum PIM Power in a Pure PIM configuration. Similarly, the CPU Power in the combined PIM+CPU system is lower than the maximum CPU Power.

In order to compare energy per computation between different configurations, we use the relevant $EPC$ values, computed by dividing the power of the relevant configuration by its throughput. That is:
\begin{gather} 
 {\textstyle 
  EPC_{PIM} = \frac{P_{PIM}}{TP_{PIM}};~~
  EPC_{CPU} = \frac{P_{CPU}}{TP_{CPU}};~~
  EPC_{Combined} = \frac{P_{Combined}}{TP_{Combined}}}.
 \label{eq:EPC_GOPS} 
\end{gather}  

The following example summarizes the entire power and energy story. Assume, again, the above shifted vector-add example using the same PIM and CPU parameters. In addition, we use $Ebit_{PIM}= 0.1$pJ~\cite{Mario2019} and $Ebit_{CPU}= 15$pJ~\cite{Connor2017}. The PIM Pure throughput is 160 GOPS (see Section \ref{sec:PIM_Throughput}) and the PIM Pure power is $P_{PIM} = \frac {Ebit_{PIM} \times R \times XBs}{CT} = \frac {0.1*10^{-12} \times 1024 \times 1024}{10^{-8}} = 10.5$W. The CPU Pure throughput (using $BW=1000$ Gpbs) is 20.8 (or 62.5) GOPS for 48 (or 16) bit DIO (see Section \ref{sec:CPU_Computations}). The CPU Pure Power is $P_{CPU} = Ebit_{CPU} \times {BW} = $15*10$^{-12} \times {10^{12}} = 15$W. A combined PIM+CPU system will exhibit throughput of $TP_{Combined} = 44.9$ GOPS and power $P_{Combined} = (\frac{P_{PIM}}{TP_{PIM}} + \frac{P_{CPU}}{TP_{CPU}}) \times TP_{Combined} = (\frac{10.5}{160\times 10^9} + \frac{15}{62.5\times 10^9}) \times (44.9\times 10^9) = 13.7$W.

Again, these results depend on the specific parameters in use. However, they demonstrate a case where, with PIM, not only the system throughput went up, but, at the same time, the system power decreased. When execution time and power consumption go down, energy goes down as well. In our example, $ECP_{CPU} = \frac{15}{20.8\times 10^9}= \frac{0.72}{10^9}$ {\small J/OP} $ = 0.72$ {\small J/GOP}, and $ECP_{Combined} = \frac{13.7}{44.9\times 10^9} = \frac{0.31}{10^9}$ {\small J/OP} $ = 0.31$ {\small J/GOP}.

\subsection{Power-Constrained Operation} 
\label{sec:Constrained_Power}

Occasionally, a system, or its components, may be power-constrained. For example, using too many XBs in parallel, or fully utilizing the memory bus may exceed the maximum allowed system or component \textit{thermal design power}\footnote{https://en.wikipedia.org/wiki/Thermal\_design\_power} ($TDP$). 
For example, the PIM power $P_{PIM}$ must never exceed $TDP_{PIM}$. When a system or a component exceeds its $TDP$, it has to be slowed down to reduce its throughput and hence, its power consumption. For example, a PIM system throughput can be reduced by activating fewer XBs or rows in each cycle, increasing the cycle time, or a combination of both. CPU power can be reduced by forcing idle time on the memory bus to limit its bandwidth (\textit{i.e.}, \textit{"throttling"}).
%
\section{The Bitlet Model - Putting it All Together}
\label{sec:Bitlet}
%

So far, we have established the main principles of the PIM and CPU performance. In this section, we first present the Bitlet model itself, basically summarizing the relevant parameters and equations to compute the PIM, CPU, and combined performance in terms of throughput. Then, we demonstrate the application of the model to evaluate the potential benefit of PIM for various use cases. We conclude with a sensitivity analysis studying the interplay and impact of the various parameters on the PIM and CPU performance and power.

\subsection{The Bitlet Model Implementation} 
\label{sec:Bitlet_Model} 



The Bitlet model consists of ten parameters and nine equations that define the throughput, power, and energy of the different model configurations. Table~\ref{tab:parameters} summarizes all Bitlet model parameters. Table \ref{tab:equations} lists all nine Bitlet equations.

PIM performance is captured by six parameters: $OC$, $PAC$, $CC$, $XBs$, $R$ and $CT$. Note that $OC$ and $PAC$ are just auxiliary parameters used to compute $CC$. CPU performance is captured by two parameters: $BW$ and $DIO$. PIM and CPU energy are captured by the $Ebit_{PIM}$ and the $Ebit_{CPU}$ parameters. For conceptual clarity and to aid our analysis, we designate three parameter types: \textit{technological}, \textit{architectural}, and \textit{algorithmic}. Typical values or ranges for the different parameters are also listed in Table~\ref{tab:parameters}. \revision{The table contains references for the typical values of the technological parameters $CT$, $Ebit_{PIM}$, and $Ebit_{CPU}$, which are occasionally deemed controversial. The model itself is very flexible, it accepts a wide range of values for \emph{all} the parameters. These values do not even need to be implementable and can differ from the parameters' typical values or ranges. This flexibility allows limit-studies by modeling systems using extreme configurations.}

The nine Bitlet model equations determine the PIM, CPU and the combined performance ($TP_{PIM}$, $TP_{CPU}$, $TP_{Combined}$), power ($P_{PIM}$, $P_{CPU}$, $P_{Combined}$), and energy per computation ($EPC_{PIM}$, $EPC_{CPU}$, $EPC_{Combined}$).

 \begin{table}[!t]
\centering 
\caption{\textit{Bitlet Model Parameters.}}
\vspace{-3mm}
\label{tab:parameters} 
\begin{tabular}{ |>{\centering\arraybackslash}m{5.5cm}|>{\centering\arraybackslash}m{2cm}|>{\centering\arraybackslash}m{2.8cm}|>{\centering\arraybackslash}m{2cm}|}
 \hline 
 {\bf Parameter name} &  {\bf Notation} & {\bf \revision{Typical Value(s)}} & {\bf Type} \\ \hline \hline 
 {PIM operation complexity} & $OC$ & 1 - \revision{64k} cycles & Algorithmic \\ \hline
  PIM placement and alignment complexity &  $PAC$  &   0 - \revision{64k} cycles & Algorithmic \\ \hline 
 {PIM computational complexity} & $CC=OC+PAC$ & 1 - 64k cycles & Algorithmic    \\ \hline  
 {PIM cycle time} &  $CT$ & 10 ns~\cite{Mario2019} & Technological \\  \hline 
 {PIM array dimensions (rows~$\times$~columns)} &  \textbf{$R \times C$} & \revision{16x16 - }1024x1024 & Technological      \\ \hline 
 {PIM array count } &  $XBs$  &   \revision{1 - 64k} & Architectural    \\ \hline 
 PIM energy for operation ($OC$=1) per bit & $Ebit_{PIM}$ & 0.1pJ~\cite{Mario2019}& Technological \\ \hline \hline
CPU memory bandwidth & $BW$ & \revision{0.1 -} 16 Tbps & Architectural \\ \hline  
  CPU data in-out bits & $DIO$ & \revision{1 - 256} bits & Algorithmic  \\ \hline
   CPU energy per bit transfer  & $Ebit_{CPU}$ & 15pJ~\cite{Connor2017} & Technological \\ \hline 
\end{tabular}
\end{table}

\renewcommand{\arraystretch}{1.5}
 \begin{table}[!t]
\centering 
\caption{\textit{Bitlet model Equations}}
\vspace{-3mm}
\label{tab:equations} 
\begin{tabular}{ |>{\centering\arraybackslash}m{5cm}|>{\centering\arraybackslash}m{6cm}|>{\centering\arraybackslash}m{1cm}|}
 \hline 
 {\bf Entity} &  {\bf Equation} & {\bf Units} \\ \hline \hline 
 {PIM Throughput} & 
 $TP_{PIM}=\frac{R \times XBs}{CC\times CT}$ &
 \small GOPS \\
 \hline  
 {CPU Throughput} & 
 $TP_{CPU}=\frac{BW}{DIO}$ & \small GOPS \\
 \hline  
 {Combined Throughput} &
 $TP_{Combined}= \frac{1}{\frac{1}{TP_{PIM}} + \frac{1}{TP_{CPU}}}$  & \small GOPS \\ 
 \hline  
 {PIM Power} & 
 $P_{PIM} = \frac {Ebit_{PIM} \times R \times XBs}{CT}$ & \small Watts\\
 \hline  
 {CPU Power} & 
 $P_{CPU} = Ebit_{CPU} \times {BW}$ & \small Watts\\
 \hline  
 {Combined Power} &
 $P_{Combined}= (\frac{P_{PIM}}{TP_{PIM}} + \frac{P_{CPU}}{TP_{CPU}}) \times TP_{Combined}$ & \small Watts\\ 
 \hline  
 {PIM Energy per Computation} & 
 \small $EPC_{PIM} = \frac{P_{PIM}}{TP_{PIM}}$ & \small J/GOP \\
 \hline 
 {CPU Energy per Computation} & 
 \small $EPC_{CPU} = \frac{P_{CPU}}{TP_{CPU}}$ & \small J/GOP \\
 \hline 
 {Combined Energy per Computation} & 
 \small $EPC_{Combined} = \frac{P_{Combined}}{TP_{Combined}}$ & \small J/GOP \\
 \hline 

 \hline  
\end{tabular}
\end{table}
\renewcommand{\arraystretch}{1}

\subsection{Applying The Bitlet Model} 
\label{sec:Applying_Bitlet_Model} 

\begin{figure}[t]
\centering
\includegraphics[width=5.9in]{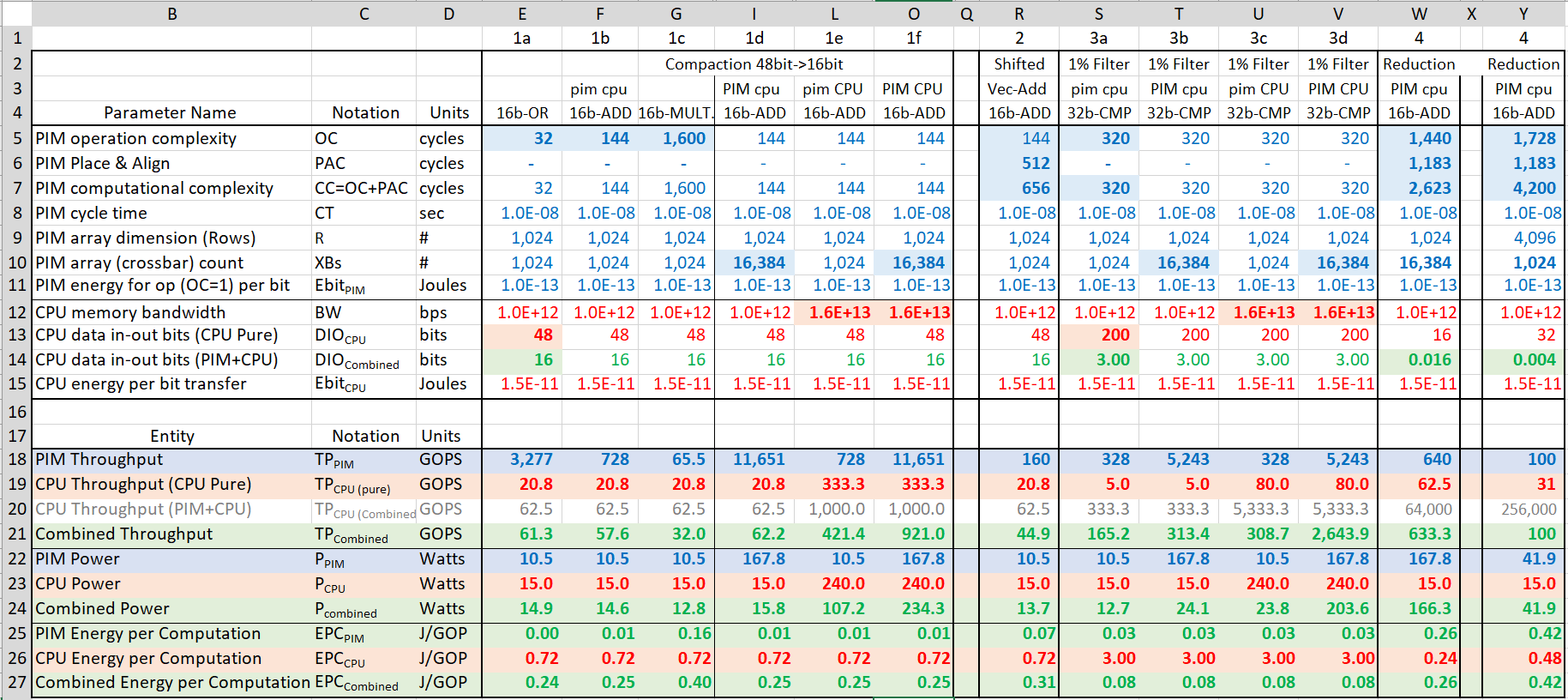}
\vspace{+0cm} 
\caption{\small {Throughput and Power comparison of CPU Pure vs. combined PIM+CPU system.}}
\label{fig:TPT-POWER-table}
\end{figure}

The core Bitlet model is implemented as a straightforward Excel spreadsheet\footnote{The spreadsheet is available at \url{https://asic2.group/tools/architecture-tools/}}. All parameters are inserted by the user and the equations are automatically computed. Figure~\ref{fig:TPT-POWER-table} is a snapshot of a portion of the Bitlet Excel spreadsheet that reflects several selected configurations.

Few general notes:
\begin{itemize} 
\item The spreadsheet can include many configurations, one per column, simultaneously, allowing a wide view of potential options to ease comparison.
\item For convenience, in each column, the model computes the three related PIM Pure (PIM), CPU Pure (CPU), and the Combined configurations. To support this, the two DIO parameters are needed; one, $DIO_{CPU}$, for the CPU Pure system, and one (usually lower), $DIO_{Combined}$, for the combined PIM+CPU system. See rows 13-14 in the spreadsheet.
\item Determining the $OC$, $PAC$, and $DIO$ parameters needs special attention. Sections \ref{sec:PIM_Computation_Principles} and \ref{sec:CPU_Computations} detail how to determine these parameters.
\item Fonts and background are colored based on the system they represent: blue for PIM, green for CPU, and red for combined PIM+CPU system.
\item Bold parameter cells with a light background mark items highlighted in the following discussions and are not inherent to the model.
\end{itemize}

Following is an in-depth dive into the various selected configurations.

\textbf{Compaction.} Cases 1a-1f (columns E-O) describe simple parallel aligned operations. In all these cases, the PIM performs a 16-bit binary computation in order to reduce data transfer between the memory and the CPU from 48 bits to 16 bits. The various cases differ in the operation type (OR/ADD/MULTIPLY, columns E-G), the PIM array count (1024/16384 XBs), and the CPU memory bandwidth (1000/16000 Gpbs) see cases 1b, 1d-1f, rows 4, 10 and 12. Note that in row 3, \textit{"pim"} means a small PIM system (1024 XBs) while \textit{"PIM"} mean a large PIM system (16384 XBs). Same holds for \textit{"cpu"} (1Tbs) and \textit{"CPU"} (16Tbs). In each configuration, we are primarily interested in the difference between the CPU and the combined PIM+CPU system results. Several observations: 

\begin{itemize} 
\item A lower $OC$ (row 5) yields higher PIM throughput and combined PIM+CPU system throughput. The combined PIM+CPU system provides a significant benefit over CPU for OR and ADD operations, yet almost no benefit for MULTIPLY.
\item When the combined throughput is close to the throughput of one of its components, increasing the other component has limited value (\textit{e.g.}, in case 1d, using more XBs (beyond 1024) has almost no impact on the combined throughput (61 GOPS), as the maximum possible throughput with the current bandwidth (1000 Gbps) is 62 GOPS).
\item When the throughput goes up, so does the power. Using more XBs or higher bandwidth may require higher power than the system $TDP$. Power consumption of over 200 Watts is likely too high. Such a system has to be slowed down by activating fewer XBs, enforcing idle time, etc...   
\item A comparison of the PIM throughput and the CPU throughput (row 18 and 20) provides a hint as to how to speed up the system. Looking at case 1b (column F), the PIM throughput is 728 GOPS while the CPU throughput is 63 GOPS. In this case, it makes more sense to increase the CPU throughput, and indeed, case 1e (row 21, column L), which increases the CPU bandwidth, improves the throughput more than case 1d (column I) does.
\end{itemize}

\textbf{Shifted Vector-add.} Case 2 (column R) summarizes the example that was widely used in Sections \ref{sec:PIM_Throughput} , \ref{sec:Combined_Throughput} and \ref{sec:Combined_Power}.

\textbf{Filter.} Case 3a-3d (columns S-V) repeats the example in Section~\ref{sec:CPU_Computations}. It describes a filter that eventually selects 1\% of the records and passes a bit-vector to identify the selected items. Each record is of $s=200$ bits. As in the compaction case above, the four configurations differ in their memory array size $XBs$ and the memory $BW$. Similar to the compaction case, we can get an idea of how to speed up the system by looking at rows 18 and 20. In this case, the PIM throughput is lower and it makes sense to add more XBs and not memory $BW$. Indeed, case 3b (column T) with stronger PIM, exhibits higher throughput than case 3c (column U) with higher memory $BW$.

\textbf{Reduction.} Case 4 (column W) reflects summing all elements in a 16-bit vector. For simplicity, we use the per-XB reduction method, $Reduction_1$, where all initial per-XB results are transferred to the CPU. On the CPU side, this computation is like a filter where only one element per XB ($p=1/R$) is transferred, and there is no need to transfer a bit-vector. On the PIM side, $CC$ is determined as described in Table~\ref{tab:PIMcycles}. With $R=1024$ rows, the number of phases is $ph=\lceil \log_2(1024) \rceil=10$. Overall, the $CC$ of the reduction is relatively high, therefore, a PIM-based reduction solution requires many XBs to be more beneficial than a Pure CPU solution.

\subsection{Impact and Interplay among Model Parameters}
\label{sec:Interplay} 

In the previous sub-section, we showed how to determine the throughput and the power of a given system configuration. Now we want to illustrate the sensitivity of the throughput and power to changes in different parameters. In this discussion, due to limited space and limited ability to visualize many parameters concurrently, we focus on the algorithmic and the architectural parameters only, \textit{i.e.}, $CC$ and $XBs$ on the PIM side and $DIO$ and $BW$ on the CPU side. The model itself, as illustrated in Figure \ref{fig:TPT-POWER-table}, supports manipulation of all parameters.

\begin{figure}[t]
\centering
\includegraphics[width=5.0in]{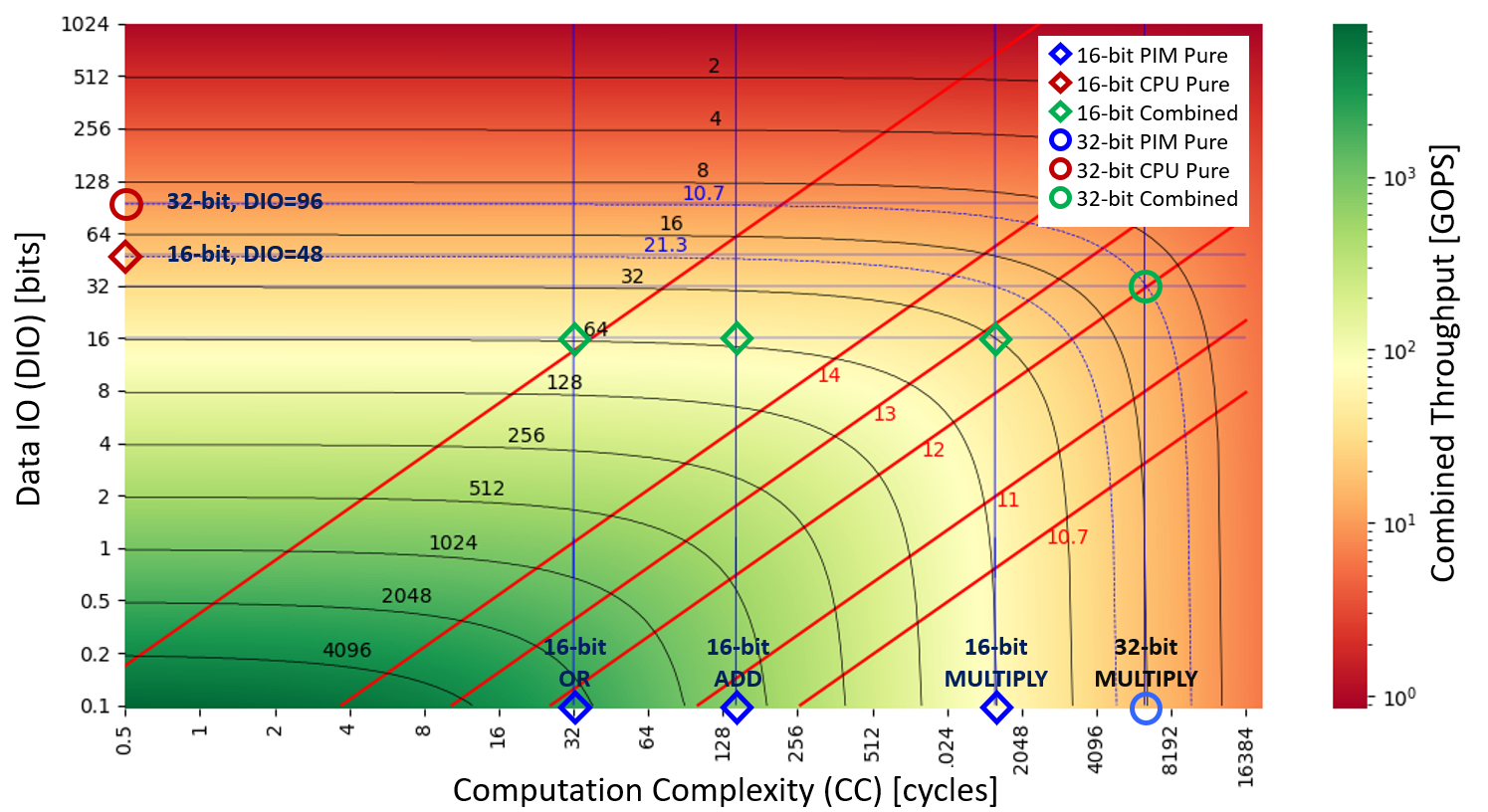}
\caption{\small {Combined Throughput [GOPS] and Power [Watt] as function of $CC$ and $DIO$. Black curved lines are equal throughput lines. Red curved lines are equal power lines. Blue horizontal (vertical) lines are equal $DIO$ ($CC$) lines to allow comparison between PIM, CPU and Combined throughput. Diamond and circle marks legend appears on the top right corner of the graph.\\
Fixed Parameters: $XBs=1024$~arrays,~$R=1024$~rows,~$BW=1000$~Gpbs, $CT=10$ns,~$Ebit_{PIM}=0.1$pJ,~$Ebit_{CPU}=15$pJ.}}
\label{fig:TPT-Fixed-XBs+BW}
\vspace{-3mm}
\end{figure}

First, in Figure \ref{fig:TPT-Fixed-XBs+BW}, we present the PIM, the CPU, and the combined PIM+CPU throughput and power as a function of $CC$ and $DIO$ for a certain PIM and CPU configuration, where $XBs=1024$ and $BW=1000$ Gbps. 
The color on the graph at point $(x,y)$ indicates the combined PIM+CPU throughput value when PIM $CC=x$ and CPU $DIO=y$. The black curved lines on the graph are equal throughput lines and are annotated with the throughput value in Gpbs. Points below a line have higher throughput and vice-versa.
The throughput value at the point ($CC=x$, $DIO=0$) reflects the PIM Pure throughput when $CC=x$. The throughput value at the point ($CC=0$, $DIO=y$) reflects the CPU Pure throughput when $DIO=y$. Note that since the axes are in logarithmic scales, the points where $CC=0$ or $DIO=0$ do not appear on the graph but can be approximated by looking at the value of the equal throughput line that is close to it. For example, the PIM Pure throughput for $CC=200$ is approximately 512 GOPS.
Blue horizontal (vertical) lines are equal $DIO$ ($CC$) lines to allow comparison between PIM, CPU, and Combined throughput.
Red curved lines on the graph are equal power lines; in this graph, power goes up when going from bottom right to top left. 
Several points were highlighted on the graph. 
Diamond-shaped points represent the three 16-bit operations (OR/ADD/MULTIPLY) mentioned in the previous sub-section (cases 1a-1c in Figure \ref{fig:TPT-POWER-table}). The circle-shaped points represent the 32-bit MULTIPLY operation. The relevant operation is marked above the relevant diamond on the $X$ axis. Observing marks with the same shape and operation type allows throughput comparison of the same operation between all three PIM, CPU, and combined PIM+CPU configurations.
Observations:


\begin{itemize} 
\item For the same $DIO$, higher $CC$ implies lower throughput.\\
With higher $CC$, PIM benefits decline, \textit{e.g.}, PIM 32/64 bit MULTIPLY has same/lower throughput than CPU. 
\item For the same $CC$, higher $DIO$ implies lower throughput.
\item An equal throughput (black) line has a knee. On the left of the knee, the throughput is impacted mostly by $DIO$, that is, the CPU is the bottleneck in this region. Below the knee, the throughput is impacted mostly by $CC$, \textit{i.e.}, the PIM is the bottleneck in this region.
\item For a given case, it is worth looking at three points: \\
(1) $x=CC,~y=0$, representing the PIM Pure throughput, \\
(2) $x=0,~y=DIO_{CPU}$, representing the CPU Pure throughput,\\
(3) $x=CC,~Y=DIO_{PIM}$, representing the combined PIM+CPU system throughput. 
\item The equal power (red) lines reveal three power regions. The top left reflects the CPU bottle-necked region, where the power is very close to CPU Pure power. The bottom right reflects the PIM bottle-necked region, where the power is very close to PIM Pure power. The power changes mainly around the knee, where each small move to the left changes the power closer to the CPU Pure power and, similarly, each small move down changes the power closer to the PIM Pure power. In the current configuration ($XBs=1024$ , $BW=1000$~Gpbs ), where CPU Pure power is higher than PIM Pure power, left means higher power and down means lower power. Different configurations may exhibit different behaviors. 
\item The linear behavior of the power lines reflects the fact that the combined PIM+CPU system power is a linear combination of the PIM Pure and CPU Pure power, where each is weighted according to the share of time they are active. When we multiply $CC$ and $DIO$ by the same number, the time ratio remains the same, and so does the combined power.
\end{itemize}

Table \ref{tab:Throughput_Examples} lists the marked points in the graph. The 64-bit MULTIPLY was added to the table to highlight a high computation complexity case where CPU Pure performs better than the combined PIM+CPU configuration.

 \begin{table}[h]
\centering 
\caption{\textit{Throughput of Binary-Operations Examples}}
\vspace{-3mm}
\label{tab:Throughput_Examples} 
\begin{tabular}{ |>{\centering\arraybackslash}m{4cm} | >{\centering\arraybackslash}m{1cm} | >{\centering\arraybackslash}m{1cm} |>{\centering\arraybackslash}m{2cm} |>{\centering\arraybackslash}m{2cm} | >{\centering\arraybackslash}m{2cm} |}
 \hline 
 {Operation} & {\bf 16-bit OR}  &  {\bf 16-bit ADD}&  {\bf 16-bit MULTIPLY}  &  {\bf 32-bit MULTIPLY}&  {\bf 64-bit MULTIPLY} \\  \hline \hline 
 {CC [cycles]} & 32 & 144 & 1600 & 6400 & 25600 \\ \hline 
 {DIO CPU / Combined [bits]} & 48 / 16 & 48 / 16 & 48 / 16 & 96 / 32 & 192 / 64 \\ \hline 
 {PIM Throughput [GOPS]} & 3277 & 728 & 65.5 & 16.4 & 4.1 \\ \hline
 {CPU Throughput [GOPS]} & 20.8 & 20.8 & 20.8 & 10.4 & 5.2 \\ \hline
 {Combined Throughput [GOPS]} & 61.3 & 57.6 & 32.0 & 10.7 & 3.2 \\ \hline
 \multicolumn{1}{|c|}{PIM Power [Watts]} & \multicolumn{5}{c|}{10.5}  \\ \hline
 \multicolumn{1}{|c|}{CPU Power [Watts]} & \multicolumn{5}{c|}{15.0}  \\ \hline
 {Combined Power [Watts]} & 14.9 & 14.6 & 12.8 & 12 & 11.4 \\ \hline 
\end{tabular}
\end{table}


\begin{figure}[t]
\centering
\includegraphics[width=5.0in]{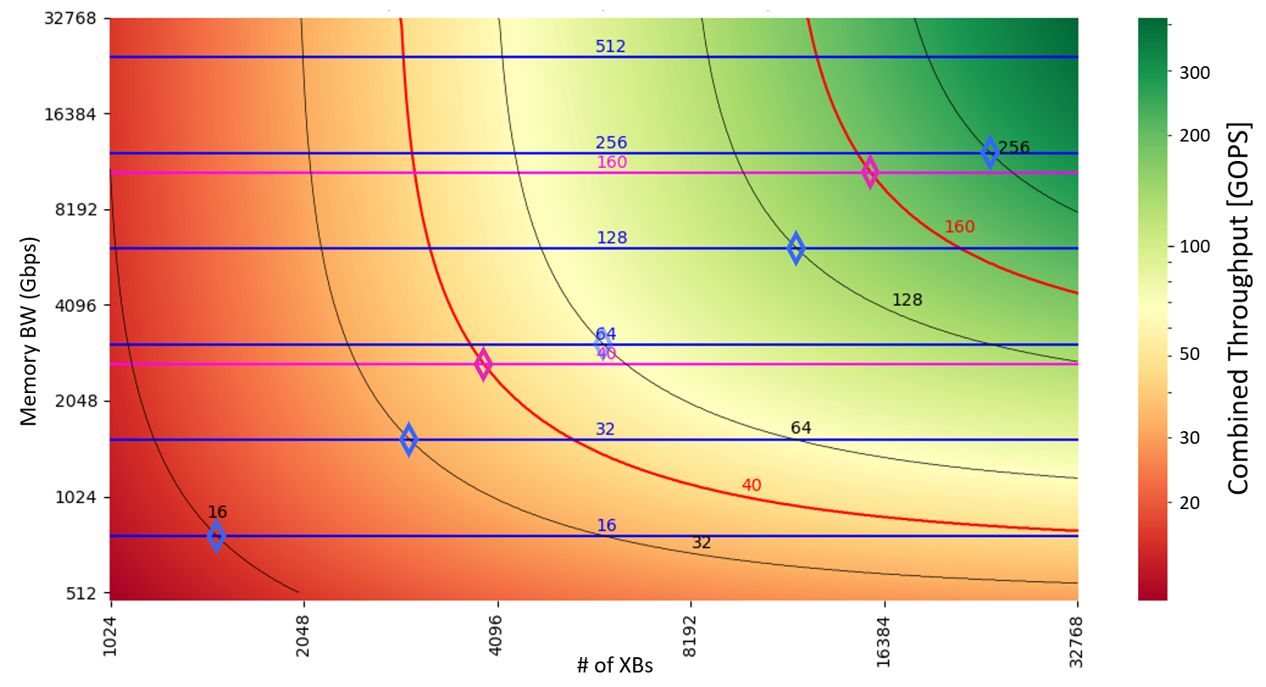}
\caption{\small {Combined throughput [GOPS] and power [Watt] as function of number of $XBs$ and memory $BW$. 
Black curved lines are equal combined throughput lines. 
Red curved lines are equal combined power lines, both at $DIO=16$ bits. 
Blue/magenta horizontal lines are equal CPU throughput/power lines at $DIO=48$ bits. Diamond marks indicate crossover point where CPU throughput (power) equals combined throughput (power).
\\Fixed parameters: $CC=6400$~cycles,~$R=1024$~rows,~$DIO=16/48$~bits, $CT=10$ns,~$Ebit_{PIM}=0.1$pJ,~$Ebit_{CPU}=15$pJ.}}
\label{fig:TPT-Fixed-CC+DIO}
\vspace{-3mm}
\end{figure}

Finally, Figure \ref{fig:TPT-Fixed-CC+DIO} presents the impact of $XBs$ and $BW$ on throughput and power. This figure assumes a certain pre-defined $CC$=6400 bits, $DIO_{Combined}$=16 bits, and $DIO_{CPU}$=48bits. The color on the graph at point $(x,y)$ indicates the combined PIM+CPU throughput value when PIM $XBs=x$ and CPU $BW=y$. The figure has curved and horizontal equal throughput and power lines. Black curved equal throughput lines and magenta curved equal power lines reflect the combined PIM+CPU configuration using $DIO$=16 bits. Blue horizontal lines reflect the CPU Pure throughput, Magenta horizontal lines reflect the CPU Pure power, both at $DIO_{CPU}$=48 bits. The diamond marks indicate throughput and power crossover points between natural trade-offs of Pure CPU at $DIO$=48 bits and combined PIM+CPU at $DIO$=16 bits. Observations:
\begin{itemize} 
\item Both throughput and power increase linearly when either $XBs$ or $BW$ increase. The crossover points help compare the CPU Pure and combined PIM+CPU alternatives. When $BW$ is high and $XBs$ is low, the PIM becomes the bottleneck and using CPU Pure is more beneficial than using PIM. On the other hand, when $BW$ is low and $XBs$ is high, the combined PIM+CPU configuration is better.  
\item The choice of working points out of all points on the same equal throughput or power line depends on the available technology and possible configurations. Bandwidth, memory size, or power limitation leaves only part of the space available, \textit{e.g.}, if $BW$ is limited to 4000 Gbps and memory size is limited to 8192K $XBs$, then roughly only points from the bottom left quarter of the graph are valid. If we also limit the power to 40 Watts, another part of the space becomes invalid.
\item We can model a PIM Pure system, by adding vertical lines to reflect PIM Pure power and performance (lines are not shown). 
\end{itemize}

The above are just several examples of the types of analyses that are enabled by the Bitlet model. The model enables analytic exploration of many parameter combinations of PIM and CPU systems.   

\subsection{Analysis of Real-Life Examples using the Bitlet Model}
\label{sec:Real_Examples} 

In the previous sections, we analyzed relatively simple and synthetic examples. In this subsection, we apply the Bitlet model on several real-life examples taken from two PIM-related papers, the Fixed Point Dot Product (FiPDP), the Hadamard Product, and the Image Convolution from the IMAGING paper~\cite{Haj2018} and the Floating Point multiplication and addition from the FloatPIM paper~\cite{FloatPIM}. All examples map useful and important algorithms onto a MAGIC-based PIM system.

In all examples, the authors made an admirable effort to compute the latency and tried to assess the throughput and power of these computations. In most cases, the authors used a single configuration ($XBs$, $R$), assumed a single value for the technological parameters, and deduced the throughput, power, and energy based on the single values.

The Bitlet model complements the above works nicely. Using their values for $CC$, the model can easily illustrate the throughput, power, and energy for different parameters. The model can help compare the results to the CPU Pure and the combined PIM+CPU systems.

\subsubsection{\bf IMAGING}
\label{sec:IMAGING} 

The IMAGING paper~\cite{Haj2018} implements several algorithms and analyzes them, but does not consider technological parameters, \textit{e.g.}, cycle time and energy, in the analysis. As a consequence, it presents results in throughput per cycle and determines area based on the number of memory cells. 
\begin{itemize}
\item \textbf{Fixed Point Dot Product (FiPDP)\footnote{https://en.wikipedia.org/wiki/Dot\_product}.} \textit{FiPDP} is a classical dot-product $S = \sum_{i=1}^{N} A_i \times B_i$, where two vectors are multiplied element-wise, and the result vector is summed. Assume two 8-bit vectors producing 16-bit interim results that are summed into 32-bit numbers. The paper assumes $R=512$. The $CC$ \revision{of the algorithm, as implemented in the IMAGING paper,} consists of the multiplication step ($12.5\times 8^2 = 800$~cycles) followed by the \revision{tree-like} reduction step ($ph \times (OC+W)+R-1 = 9\times (288+32) +511 \approx 3391$ cycles). The two steps together take approximately $4200$~cycles. The paper neither states the throughput of this operation nor does it make any sensitivity analysis. Using the Bitlet model, we can easily compute the throughput and analyze its sensitivity. For example, for $XBs=512$, $R=512$, and $CT=10ns$, we achieve PIM Pure and combined PIM+CPU throughput of about $6$ GOPS, which is rather low compared to the CPU Pure throughput of $31$ GOPs at $BW=1000$~Gpbs. Using a configuration of $XBs=4096$ and $R=1024$ increases the PIM Pure (and combined PIM+CPU) throughput to about $100$~GOPS, which is higher than the CPU Pure throughput of $31$ GOPS stated above. 

\item \textbf{Hadamard Product\footnote{https://en.wikipedia.org/wiki/Hadamard\_product\_(matrices)}.} The \textit{Hadamard Product} is an element-wise $K\times K$ matrix product, that is, for all $(i,j)$, $C_{i,j} = A_{i,j} \times B_{i,j}$. In fact, it is equivalent to an element-wise $N$ vector product, that is, for all $i$, $C_{i} = A_{i} \times B_{i}$.  The paper focuses on 8-bit pixels as the elements to multiply. If memory space is scarce, several pairs of elements are located in the same row to fit the matrices in the available memory. 
The paper also considers the case where the input vectors are larger than the size of available memory, so the computation needs to be repeated. None of these manipulations affect the computation throughput since they result in doing, \textit{e.g.}, $10\times$ more work in $10\times$ longer time. For throughput computation, we use the Bitlet model assuming a single multiplication in each row. Doing so, we can obtain the real throughput in GOPS and compare it to the CPU Pure and the combined PIM+CPU system throughput. 

Table \ref{tab:Hadamard} provides several examples with varying $XBs$ and $R$ values. We assume the paper's original value of $CC=710$ cycles. We use the Bitlet model to also compute the PIM Pure, CPU Pure, and combined PIM+CPU system throughput. For the CPU configuration, we assume $BW=1000$ Gbps, $DIO=32$ bits for CPU Pure, and $DIO=16$ bits for the combined system. As expected, the throughput goes up with the number of $XBs$ (and $R$). For low numbers of $XBs$ and $R$, CPU Pure is better than a combined PIM+CPU system ($31$ GOPS vs. $23$ GOPS). However, adding XBs improves the combined PIM+CPU system throughput compared to CPU Pure, providing over $49$ GOPS vs. $31$ GOPS, respectively.

 \begin{table}[h]
\centering 
\caption{\textit{Throughput of the Hadamard Product}}
\vspace{-3mm}
\label{tab:Hadamard} 
\begin{tabular}{|>{\centering\arraybackslash}m{1cm}|>{\centering\arraybackslash}m{1cm}|>{\centering\arraybackslash}m{1.5cm}|>{\centering\arraybackslash}m{2cm}|>{\centering\arraybackslash}m{1.5cm}|>{\centering\arraybackslash}m{1.5cm}|>{\centering\arraybackslash}m{2cm}|}
 \hline 
{\bf $\bm{XBs}$}  &  {\bf $\bm{R}$}  &  {\bf $\bm{CC}$ \small[Cycles]}  &  {\bf $\bm{TP_{/cycle}}$ \small[GOP/Cycle]}&  {\bf $\bm{TP_{PIM}}$ \small[GOPS]}& {\bf $\bm{TP_{CPU}}$ \small[GOPS]}& {\bf $\bm{TP_{Combined}}$ \small[GOPS]} \\  \hline \hline 
512 & 512 & 710 & 369 &  37  & 31 & 23 \\ \hline
1,024 & 512 & 710 & 738 &  74  & 31 & 34 \\ \hline
4,096 & 1,024 & 710 & 5,907 &  591  & 31 & 57 \\ \hline
16,384 & 1,024 & 710 & 23,630 &  2,363  & 31 & 61 \\ \hline
\end{tabular}
\end{table}

\item \textbf{Image Convolution\footnote{https://en.wikipedia.org/wiki/Kernel\_(image\_processing)}.} A single convolution computation consists of multiplying a $P \times P$ pixel window in a picture with a $P \times P$ coefficient matrix and creating a new picture where the center element of the selected window is set with the newly computed value. $P$ is usually a small odd number, \textit{e.g}., 3 or 5. Each pixel is $W$-bits wide. The IMAGING paper goes a long (and smart) way to implement the convolution on top of the MAGIC NOR memory array and computing its latency. For our discussion, we need to consider only the following: (a) Computing each pixel involves $P^2$ $W$-bit multiplications, ($P^2-1$) $2W$-bit additions, $W\times P \times (P-1)$ HCOPY operations and ($P-1$) VCOPY operations. The last $\frac{P-1}{2}$ pixels in each row (\textit{e.g.}, 1 or 2 pixels for $P=3,~5$) are duplicated at the beginning of the next row. Therefore, to reduce space overhead, each row has to have a minimal number of pixels. In the aforementioned examples, $8+1=9$ pixels per row for $P=3$ and $8+2=10$ for $P=5$. Table \ref{tab:Convolution1} lists the $CC$ for convolutions using $W=8$ bits, $P=3,~5$ and $R=512,~1024$. The table clearly shows that convolution has a very high computation complexity. 

\begin{table}[h]
\centering 
\caption{\textit{Convolution Computation Complexity.}}
\vspace{-3mm}
\label{tab:Convolution1} 
 \begin{tabular}{|>{\centering\arraybackslash}m{0.5cm}|>{\centering\arraybackslash}m{2cm}|>{\centering\arraybackslash}m{2cm}|}
\hline 
{\bf $\bm{P}$}  &  {\bf $\bm{CC}$ ($\bm{R=512}$) [Cycles]}  &  {\bf $\bm{CC}$ ($\bm{R=1024}$) [Cycles]} \\  \hline \hline 
3 &  69,296  &  77,488 \\ \hline
5 &  188,592  &  204,976 \\ \hline
\end{tabular}
\end{table}

At this point, we can use the Bitlet model and obtain the throughput values. One immediate observation is that since the input and output matrices have the same size, there is no data transfer reduction, and the value of using PIM as a pre-processing stage is questionable. In other words, both $DIO_{CPU}$=16 and $DIO_{Combined}$=16, thus the CPU Pure throughput is higher than that of the combined PIM+CPU throughput. We ignore this concern and compare PIM Pure to CPU Pure.
Results are shown in Table \ref{tab:Convolution2}. The table shows that convolution is significantly heavier than the previous examples examined above. This is expected, as convolution involves many multiplications per pixel, especially when $P=5$. According to the model, only a huge PIM configuration ($XBs=64K, R=1024$) may compete with CPU Pure. One may question even that, since the power needed for this configuration, obtained from the Bitlet model as well, is approximately $650$ Watts. It is worth noting that this computation is quite heavy for the CPU core as well. Every pixel requires (for $P=3$) $3^2$=9 8-bit multiplications and $3^2$-1=8 16-bit additions, so to sustain $63G$ convolutions per second, the CPU needs to perform about $63G \times 17 \simeq 1T$ instructions per second. Achieving that throughput requires, for example, four 4-GHz high end CPUs, supporting two wide SIMD instructions (\textit{e.g.}, \textit{AVX-512}\footnote{https://en.wikipedia.org/wiki/AVX-512}) per cycle. 

\begin{table}[h]
\centering 
\caption{\textit{Convolution Throughput.}}
\vspace{-3mm}
\label{tab:Convolution2} 
\begin{tabular}{|>{\centering\arraybackslash}m{0.5cm}|>{\centering\arraybackslash}m{1cm}|>{\centering\arraybackslash}m{1cm}|>{\centering\arraybackslash}m{1.5cm}|>{\centering\arraybackslash}m{2cm}|>{\centering\arraybackslash}m{1.5cm}|>{\centering\arraybackslash}m{1.5cm}|>{\centering\arraybackslash}m{2cm}|}
\hline 
{\bf $\bm{P}$}  &{\bf $\bm{XBs}$}  &  {\bf $\bm{R}$}  &  {\bf $\bm{CC}$ \small[Cycles]}  &  {\bf $\bm{TP_{/cycle}}$ \small[GOP/Cycle]}&  {\bf $\bm{TP_{PIM}}$ \small[GOPS]}& {\bf $\bm{TP_{CPU}}$ \small[GOPS]}& {\bf $\bm{TP_{Combined}}$ \small[GOPS]} \\  \hline \hline 
3 &  1,024  &  1,024  &  77,488  & 14 &  1.4  & 63 & 1.3 \\ \hline
3 &  8,192  &  1,024  &  77,488  & 108 &  10.8  & 63 & 9.2 \\ \hline
3 &  65,536  &  1,024  &  77,488  & 866 &  86.6  & 63 & 36.3 \\ \hline
5 &  1,024  &  1,024  &  204,976  & 5 &  0.5  & 63 & 0.5 \\ \hline
5 &  8,192  &  1,024  &  204,976  & 41 &  4.1  & 63 & 3.8 \\ \hline
5 &  65,536  &  1,024  &  204,976  & 327 &  32.7  & 63 & 21.5 \\ \hline
\end{tabular}
\end{table}
\end{itemize}

\subsubsection{\bf FloatPIM}
\label{sec:FloatPIM} 

The FloatPIM paper implements a fully-digital
scalable PIM architecture that natively supports floating-point operations. As opposed to the IMAGING paper, FloatPIM does address time and power evaluation. In this section, we discuss the in-memory floating-point operation used in FloatPIM.

A floating-point multiply operation takes $T_{Mul} = 12N_e+6.5N_m^2+7.5N_m-2$ cycles, where $N_m$ and $N_e$ are the number of mantissa and exponent bits, respectively. Similarly, floating-point add  operation takes $T_{Add} = (3 + 16N_e + 19N_m + N_m^2)$ NOR cycles and $(+2N_m + 1)$ \textit{search} cycles. For simplicity, we assume here that NOR and search cycles have the same cycle time. 
The paper uses the \textit{bfloat16}\footnote{https://en.wikipedia.org/wiki/Bfloat16\_floating-point\_format} number format where $N_m=7$ and $N_e=8$. Following that, $T_{Mul}=360$ cycles and $T_{Add}=328$ cycles. On average, each of the two \textit{bfloat16} operations takes $CC \simeq 344$ cycles.

We tried to approximate the FloatPIM floating-point throughput and power using the Bitlet model. In particular, we tried to understand the sensitivity of these numbers to the technological cycle time and energy model parameters. Bitlet default parameters for $CT$ and $Ebit_{PIM}$ are 10ns and 0.1pJ, respectively. The FloatPIM equivalents are 1.1ns and 0.29fJ. Table \ref{tab:FloatPIM} shows the significant impact of the model parameters on the results. The first line in the table uses FloatPIM parameters, while the second line uses the Bitlet model default parameters. 
The results differ a lot, but once shown, seem quite obvious. A 9$\times$ faster cycle time increases the throughput by 9$\times$. Reducing energy per bit by $345\times$ increases computation per Joule by $345\times$, and, finally, accounting the two differences combined, increases power by $\frac{345}{9.1}=38\times$. Note that FloatPIM uses near-memory functions in addition to in-memory functions to implement \textit{bfloat16} add. Our comparison focuses on highlighting the impact of the model parameter setting, so we have accounted MAGIC-NOR cycles only and ignored the near-memory work. 
\begin{table}[h]
\centering 
\caption{\textit{FloatPIM parameters vs. Bitlet Defaults.}}
\vspace{-3mm}
\label{tab:FloatPIM} 
\begin{tabular}{|>{\centering\arraybackslash}m{1cm}|>{\centering\arraybackslash}m{1cm}|>{\centering\arraybackslash}m{0.5cm}|>{\centering\arraybackslash}m{1.1cm}|>{\centering\arraybackslash}m{1.2cm}|>{\centering\arraybackslash}m{1.2cm}|>{\centering\arraybackslash}m{1.6cm}|>{\centering\arraybackslash}m{1cm}|>{\centering\arraybackslash}m{1cm}|>{\centering\arraybackslash}m{1.8cm}|}
\hline 
\textbf{Model} & \textbf{$\bm{XBs}$}& \textbf{$\bm{R}$} & \textbf{$\bm{CC}$ \small [Cycles]} & \textbf{$\bm{CT}$}\newline\textbf{\small[sec]}& \textbf{$\bm{Ebit_{PIM}}$ \small [Joule]} & \textbf{$\bm{TP_{/cycle}}$ \small [GOP/Cycle]} & \textbf{$\bm{TP_{PIM}}$ \small [GOPS]} & \textbf{$\bm{P_{PIM}}$ \small [Watt]} & \textbf{$\bm{TP_{PIM}/P_{PIM}}$ \small [GOPS/Watt]} \\ \hline \hline
FloatPIM & 65,536 & 1024& 336.5 & 1.10E-09 & 2.90E-16 & 199,432 & 181,302 &  18 & 10247 \\ \hline
Default & 65,536 & 1024 & 336.5 & 1.00E-08 & 1.00E-13 & 199,432 & 19,943 & 671 & 30  \\ \hline
\end{tabular}
\end{table}
\\
Two observations from the FloatPIM analysis:
\begin{itemize}
\item The choice of \textit{bfloat16} is quite beneficial. The \textit{bfloat16} add/multiply computation complexity is $328/380$ cycles, quite reasonable compared to fixed32 add/multiply computation complexity of $288/6400$ cycles.
\item The choice of technological (and other) parameters has a major impact on the results. The $345\times$ difference in GOPS per Watt is quite significant when comparing a PIM system to a CPU system.  
\end{itemize}

\subsection{Model Limitations}
\label{sec:Model_Limitations} 

As in many models, the Bitlet model trades accuracy with simplicity. In this section, we list several model limitations that Bitlet users should be aware of. Some of these limitations will be addressed in future versions of Bitlet. The list below distinguishes between limitations due to lack of refinement and unsupported features.
 
\textbf{Potential model refinement:}
\begin{itemize}
\item \textbf{Inter-XB Copying.}  The model ignores inter-XB copying (see Section \ref{sec:PIM_Computation_Principles}). Some use cases may require many inter-XB copies, and accounting for them will improve the model accuracy. Adding inter-XB copying is challenging since it requires modeling of the memory internal busses.

\item \textbf{Impact of Arithmetic Intensity.} The model assumes that in PIM-relevant workloads, the CPU throughput is solely determined by the data transfer throughput (Section \ref{sec:CPU_Computations}). This assumption is valid for today's data-intensive applications, and it simplifies the Bitlet model tremendously. If, in the future, the memory bus bandwidth increases to the point it is no longer a bottleneck, the CPU core activity will have to be taken into consideration when assessing the CPU throughput.

\item \textbf{Cell Initialization.} Depending on the PIM logic technology and the specific basic operation in use, an output cell may need to be initialized before it is computed (\textit{e.g.}, to $R_{ON}$ in MAGIC NOR based PIM). The extra initialization cycles can potentially double the PIM execution time and should be considered in the computation complexity.

\item \textbf{Row Selection.} When computing power, the current model assumes that at every PIM cycle, all cells in the target column consume energy. This assumption may be false if row selection is used. Counting all rows instead of only the participating rows increases the energy estimate and degrades the model accuracy. This may be significant in algorithms that make serial $VCOPY$s, like shifted vector-add and reductions (see Section~\ref{sec:PIM_Computation_Principles}). 

\item \revision{\textbf{Comparing PIM to systems other than CPU.} The current Bitlet model supports PIM, CPU, and combined PIM and CPU systems. Extending the model to support other systems, \textit{e.g.}, GPU, is conceptually similar to modeling a CPU, as long as the data transfer remains the main system bottleneck. In a high level, only the non-PIM parameters $BW$, $DIO$, and $Ebit_{CPU}$ need to be modified in order to model a GPU.}

\end{itemize}

\textbf{Potential new features:}
\begin{itemize}
\item \textbf{Pipelined PIM and CPU.} So far, we assumed that PIM computation and data transfer cannot overlap. \revision{We can achieve such overlapping by employing a mechanism similar to double buffering\footnote{https://wiki.osdev.org/Double\_Buffering}. That is, we dynamically divide the available XBs into two groups. While one group performs PIM computation, the other does the memory to CPU data transfer, and vice versa. Doing so, the PIM computation may take twice the time, but data transfer can operate continuously. As a result, if the memory bus is the bottleneck (consuming more than half of the total time), the total time can be reduced from $({T_{PIM} + T_{CPU}})$ to $\max(T_{CPU}, 2 \times T_{PIM})$. The throughput (and the power) increase accordingly}. 

\item \textbf{Endurance and Lifetime.} 
\revision{Low endurance is a major obstacle for achieving a reasonable lifetime in memristor-based PIM systems, due to the high rate of memory writes when PIM is employed.} 
The current Bitlet model does not support endurance and lifetime considerations or estimates. Since the model does count the $CC$ cycles, it can help count cell writes, and hence, help in assessing endurance impact on lifetime.

\item \textbf{Non-single-row Based PIM Computations.} Bitlet assumes the single row-based computing principle, where each row contains a separate computation element and, in each cycle, all of the rows may participate in computation concurrently (Section \ref{sec:Logic_Execution}). In some PIM use cases, a record may span over more than one row to either improve latency or to locate long data elements within a short row. The model can support such cases, assuming the $CC$ and the $R$ parameters are carefully computed to reflect this.
\end{itemize} 

\section{Conclusions}
\label{sec:conclusion}

This paper motivates and describes Bitlet, a parameterized analytical model for comparison of PIM and CPU systems in terms of throughput and power. We explained the PIM computation principles, presented several use cases, and demonstrated how the model can be used to analyze real-life examples. We showed how to use the model to pinpoint when PIM is beneficial and when it is not, and to understand the related trade-offs and limits. We believe the model provides insights into how stateful logic-based PIM performs.

We analyzed several selected PIM and CPU systems and some insights following this analysis. For example, the effectiveness of a PIM system depends on several parameters, \textit{e.g.}, the degree of parallelism, data reduction potential, and power limitations of the architecture. In our analysis, we stabilized several model parameters and performed only a partial analysis of the systems, mainly for demonstration of the model abilities and features. Many more systems can be fully explored by the Bitlet model, and we expect more insights will be reached.

In the future, we plan to extend the Bitlet model and refine it to consider inter-XB copying, the impact of arithmetic intensity, cell initialization, and row selection. Such model refinements will increase the model accuracy. We also plan to add new features, \textit{e.g.}, endurance/lifetime estimation and non-single-row based PIM evaluation. The former will provide deeper inspection, analysis, and comparison of PIM systems, while the latter will significantly expand the span of PIM systems that can be analyzed by the Bitlet model.

\bibliographystyle{ACM-Reference-Format}
\bibliography{main}

\end{document}